\documentclass{aa}
\usepackage[utf8]{inputenc}
\usepackage{slashbox}
\usepackage{natbib}
\usepackage[colorlinks]{hyperref}
\usepackage{xcolor}
\usepackage{ulem}
\usepackage{booktabs}
\hypersetup{
    citecolor=[RGB]{44, 117, 255},  
    linkcolor=[RGB]{231, 62, 1},
    urlcolor=[rgb]{0,0,1}
}

\usepackage[]{color}

\definecolor{blue}{RGB}{0, 165, 255}

\definecolor{green}{RGB}{106,180,153}

\definecolor{violet}{RGB}{212,115,212}

\usepackage{ulem}

\graphicspath{{./images/}}

\makeatletter
\renewcommand*\aa@pageof{, page \thepage{} of \pageref*{LastPage}}
\makeatother

\title{Gas distribution from clusters to filaments in IllustrisTNG}

\titlerunning{}
\author{
  C.~Gouin\inst{1,2}\thanks{E-mail:~\tt{celinegouin@kias.re.kr}},  
  S.~Gallo\inst{1},
  N.~Aghanim\inst{1} 
}
\institute{
Université Paris-Saclay, CNRS, Institut d’Astrophysique Spatiale, 91405, Orsay, France
\label{inst1}
\and
School of Physics, Korea Institute for Advanced Study (KIAS), 85, Hoegiro, Dongdaemun-gu, Seoul, 02455, Republic of Korea
\label{inst2}
}
\date{\today}

\abstract{
Matter distribution in the environment of galaxy clusters, from their cores to their connected cosmic filaments, must be in principle related to the underlying cluster physics and it evolutionary state.
We aim to investigate how radial and azimuthal distribution of gas is affected by cluster environments, and how it can be related to cluster mass assembly history. \\
Radial physical properties of gas (velocity, temperature, and density) is first analysed around $415$ galaxy cluster environments from IllustrisTNG simulation at $z=0$ (TNG300-1).
Whereas hot plasma is virialised inside clusters ($< R_{200}$), the dynamics of warm hot inter-galactic medium (WHIM) can be separated in two regimes: accumulating and slowly infalling gas at cluster peripheries ($\sim R_{200}$) and fast infalling motions outside clusters ($> 1.5 R_{200}$). 

The azimuthal distribution of dark matter (DM), hot and warm gas phases is secondly statistically probed by decomposing their 2-D distribution in harmonic space.
Inside clusters, the azimuthal symmetries of DM and hot gas are well tracing cluster structural properties, such as their center offsets, substructure fractions and elliptical shapes. Beyond cluster virialised regions, we find that WHIM gas follows the azimuthal distribution of DM thus tracing cosmic filament patterns.
Azimuthal symmetries of hot and warm gas distribution are finally shown to be imprints of cluster mass assembly history, strongly correlated with the formation time, mass accretion rate, and dynamical state of clusters. \\
Azimuthal mode decomposition of 2-D gas distribution is a promising probe to assess the 3-D physical and dynamical cluster properties up to their connected cosmic filaments.
}


\authorrunning{Gouin et al.}

\begin{document}
\maketitle

\section{Introduction}

Galaxy cluster environments are ideal laboratories to probe both, the building up of massive structures, and the complex physics of baryons.
These most massive gravitationally bound objects of the Universe are located at the nodes of the underlying large-scale cosmic web \citep{deLapparent1986,Bond1996} and they link a network of cosmic filaments mainly composed of dark matter (DM), which constitutes the cosmic skeleton framework, along which baryons flow and collapse.
Under the action of gravity, large-scale cosmic flows transport matter on cosmic large-scale structures from voids to sheets, and then via elongated filaments into clusters \citep{Zeldovich1970}.
The anisotropic large-scale matter distribution and its associated accretion processes have been investigated both theoretically and via N-body simulations \citep[][]{Pichon2010,Codis2015,Shim2020}, to establish a picture of the evolutionary and dynamical aspects of the cosmic network \citep[see e.g.][for a review]{Hahn2016_review}.
Given the large diversity of filament types \citep[in term of length and width, see for example][]{Cautun2014,Galarraga2020}, different approaches have been developed to detect them, and lead to their own filament definition. 
One can cite topology-based \citep{Aragon2010_spineweb,Sousbie2011}, hessian-based \citep{Hahn200_TWEB,Cautun2013} and geometry-based \citep{Tempel2016,Pereyra2020,Bonnaire2020,Bonnaire2021} filament-finder techniques.
These cosmic web classification and detection methods are crucial to explore how cosmic web environment drives the physical properties of it content: gas, galaxies, and DM.

Focusing on the gas component, gas filaments in the cosmic web are currently challenging to detect due to e.g. low X-ray emissivity, low signal to noise ratio, background contamination, etc. Only few observations of cosmic gas filaments have been reported, such as individual massive bridge of hot gas around or between clusters \citep{Planck2013_filament,Eckert2015,Akamatsu2017,Nicastro2018,Bonjean2018}, and statistically characterised by averaging gas filament profiles using stacking techniques \citep{Tanimura2019,deGraaff2019,Tanimura2020A,Tanimura2020B}.
In order to interpret and prepare upcoming observations of gas filaments, predictions of cosmic gas properties from state-of-art hydrodynamical cosmological simulations are essential.
\cite{Martizzi2019} have shown that about 46\% of baryons is supposed to be in the form of warm hot intergalactic medium (WHIM). 
The majority of this warm gas phase is expected to be located inside filamentary structures, and must account for 80\% of the baryonic budget \citep{Galarraga2021,Galarraga2021b}.
It makes WHIM gas phase a powerful tracer of the cosmic web which should constitute a reservoir of baryons that could solve the so-called \textit{missing baryon problem} \citep{Cen1999,Dave2001_WHIM}. 
For these reasons, the gaseous component of the cosmic web is becoming the subject of more and more studies, with the aim to construct a comprehensive picture of the baryonic physical processes (heating, cooling, shocks, etc.) that gas undergoes during its transit from one cosmic environment to another \citep{Gheller2019,Martizzi2019,Tuominen2021,Zhu2021}.
This interest for cosmic gas is further enhanced by the prospect of future missions that will allow us to explore the hidden cosmic gas with unprecedented accuracy in the coming years \citep[][]{Simionescu2021}.

In this context, the outskirts of galaxy clusters constitute unique regions, where cosmic filaments intersect and are the most easily detectable due to larger density contrasts \citep[as observed via galaxy distribution around clusters][]{Mahajan2018,Einasto2020,Malavasi2020,Gouin2020}.
Cosmic gas infalls from the large-scale cosmic web to clusters tunnelled by their connected filaments.
Due to the dissipative nature of the gas component, it must undergo a large variety of complex physical mechanisms, such as accretion shocks \citep{Shi2020,Zhang2021}, dynamical interaction with the hot gas inside clusters, gas disruption of infalling galaxies \citep{Mostoghiu2021}, turbulent motions \citep{Rost2021}, etc. 
One can cite \cite{Walker2019}, for a complete review on gas cluster outskirts. 

In contrast, inside the gravitational potential wells of clusters, the intra-cluster medium (ICM) is accumulating mainly in the form of a hot plasma.
In these central regions, gas is assumed to be at the hydrostatic equilibrium, and spherically distributed inside galaxy clusters.
Nevertheless, both observations and simulations have shown that clusters are not perfectly at the hydrostatic equilibrium due to turbulence and bulk motions which arise at cluster peripheries \citep[see e.g.][]{Angelinelli2020,Ansarifard2020,Gianfagna2021}. 
In order to probe the thermodynamical state of ICM, a powerful technique is to explore the azimuthal gas distribution. Indeed, \cite{Chen2019} showed that the elliptical shape of ICM correlates with the amount of non-thermal pressure support, and can be related to the mass accretion history of clusters.
It is therefore crucial to accurately measure the anisotropy of gas distribution from the ICM to cosmic web filaments, to better constrain relations between cosmic gas distribution and cluster evolution.

In order to assess deviations from spherical symmetry of gas distribution in clusters, different techniques have been developed, such as the asymmetry parameter \citep{Schade1995}, centroid shift \citep{Mohr1993}, light concentration ratio \citep{Santos2008}, Gaussian fit parameter \citep{Cialone2018}, etc., and the combination of these morphological parameters \citep[see e.g.][]{DeLuca2020}. Recently, \cite{Capalbo2021} have also proposed to infer cluster morphology by modelling 2-D cluster gas maps with Zernike polynomials.
These various techniques are powerful to quantify the degree of disturbance in the cluster shape and are good proxies of cluster dynamical state. 
However, they remain focused on cluster morphology, and on the analysis of gas distribution in the most inner part of clusters (typically up to $R_{500}$). Beyond the virial radius, the azimuthal scatter technique was proposed to quantify departures from spherical symmetry in the radial profiles of gas properties \citep{Vazza2011}, and successfully traced the thermodynamical state of the ICM \citep{Eckert2012,Roncarelli2013,Ansarifard2020}. 
However, the azimuthal scatter method is sensitive to all kinds of deviations arsing from the gas clumping, the cluster ellipticity, and the large-scale anisotropic structures, without allowing to identify individually these different features.

Here, we propose to use an alternative technique based on 2-D decomposition of matter distribution in harmonic modes. This method has emerged to quantify separately the different azimuthal symmetries inside a given aperture centered on clusters. This aperture multipole technique has been developed for weak-lensing mass map applications \citep{Schneider1997}, and succeeded in estimating both, elliptical shape of clusters \citep[see e.g.][]{Clampitt2016B,Shin2018}, and filamentary patterns at cluster outskirts \citep{Dietrich2005,Mead2010,Gouin2017}. 

Such a multipole decomposition method can be also performed in 3-D and can be weighted by different variables such as mass, velocity, turbulence, etc. Recently, \cite{Valles2020} have used similar approach to study angular distribution of gas accretion flows in simulation. By probing gas flows in 3-D harmonic (with $l,m$ harmonic modes), they found that the overall patterns can be described by focusing only on the main contributions in the projected angle approximation (with $l=0$ and $m\neq0$).
Moreover, the 2-D harmonic decomposition has the advantage of being more easily applicable to observational data. Indeed, the 2-D multipole moment formalism has been already applied on projected photometric galaxy distribution \citep{Gouin2020} and on weak lensing maps in observation \citep[see e.g.][]{Dietrich2005}. This 2-D formalism appears efficient to probe angular features in cluster environments, both cluster elliptical shape and filamentary pattern. Using Zernike polynomials decomposition on Sunyaev-Zel'dovich (SZ) mock images of simulated clusters, \cite{Capalbo2021} have also proved the efficiently of such methods on quantifying morphology and dynamical state of clusters.

In the present work, we will apply 2-D aperture multipole moment decomposition to gas distribution inside and around clusters. By separating gas in different main phases, we aim at highlighting which gas phase preferentially traces the cluster shape and the large-scale filamentary pattern.
This method will be also used to distinguish between the different features on the gas distribution such as the amount of substructures, the halo ellipticity, and the connected cosmic filaments.
The gas azimuthal symmetries will be finally probed to investigate if the azimuthal gas distribution traces the cluster dynamics and its accretion history. 

This paper is organized as follows.
In Section 2, we describe our sample of $415$ simulated cluster extracted from IllustriTNG simulation \citep{ILLUSTRIS_TNG}, and their physical properties.
In Section 3, we start by investigating the properties of the gas as a function of the cluster-centric distance and cluster mass. This allows us to choose which gas phases and radial apertures are optimal choices to investigate further spherical symmetry deviations of gas distribution.
In Section 4, we present the multipole moment formalism. We probe different azimuthal symmetries of hot gas and DM inside clusters, and show how they are related to structural properties of cluster halo. 
In Section 5, the average level of azimuthal symmetries of hot gas, WHIM and DM are estimated inside and outside clusters. These deviations from circular symmetry are compared to cluster physical properties and to their recent mass assembly history. Finally, we discuss and summarize our conclusions in Section 6.

\section{Simulated cluster sample \label{SEC:CLUSTER}}

In this section, we present our sample of $415$ simulated cluster environments, extracted from IllustrisTNG simulation \citep{ILLUSTRIS_TNG}, for which various physical and structural properties has been previously estimated in \cite{Gouin2021}. 

\subsection{Cluster environments from IllustrisTNG simulation}

The large cosmological magneto-hydrodynamical IllustrisTNG simulations \citep{ILLUSTRIS_TNG} provide the spatial and dynamical evolution of dark matter, gas, stars, and black holes on a moving-mesh code \citep{AREPO}, and assume cosmological parameters from the \textit{Planck} 2015 results \citep{Planck2016}.
Considering the series of IllustrisTNG simulation boxes, we focus here on IllustrisTNG300-1 at $z=0$; the cubic box has a length of $302.6 \text{Mpc}$ and the mass resolution is about $m_{\text{DM}} = 4.0 \times 10^7 M_{\odot}/h$. This large and high-resolution simulation box is ideal to accurately describe matter distribution around galaxy clusters up to their large scale environments at $z=0$.

Our sample of galaxy clusters is based on the halo catalog provided by IllustrisTNG and identified with a friends-of-friends (FoF) algorithm \citep{Davis1985}. Notice that the radial physical scale $R_{200}$ of FoF halos is defined as the radius of a sphere centered on the halo which encloses a mass $M_{200}$ and a density equals $200$ times the critical density of the Universe at $z=0$.
The IllustrisTNG simulations provide also subhalo catalogues derived by the Subfind algorithm \citep{subfind}, which allow to quantify the amount of substructures inside a given host halo.
Starting from the IllustrisTNG halo catalog at $z=0$, we select all FoF halos with masses $M_{200}>5 \times 10^{13} M_{\odot}/h$ that are more distant than $5 \, R_{200}$ from the simulation box edges.
Our sample contains $415$ clusters which can be divided in two distinct mass bins: the $266$ massive groups with mass $M_{200} =[5-10] \times 10^{13} M_{\odot}/h$, and $149$ galaxy clusters with $M_{200} > 1 \times 10^{14} M_{\odot}/h$. Notice that if we do not distinguish between galaxy groups and clusters, we otherwise refer to  the $415$ most massive halos as our cluster sample.

\subsection{Physical and structural properties}

We refer the reader to \cite{Gouin2021}, for details on the computation of physical and structural properties of our cluster sample. Here, we summarise the definitions and computation procedures of the different estimated parameters.

\textbf{Mass assembly history.}
In order to probe the mass assembly history of clusters, the time evolution of cluster mass $M_{200}(z)$ has been computed using the available merger tree of subhalos computed with the SubLink algorithm \citep[see][for details on merger tree computation in Illustris]{Rodriguez2015}. Two distinct proxies of the mass assembly history of clusters have been estimated: the formation redshift $z_{form}$ and the mass accretion rate $\Gamma$. These two parameters provide complementary information to quantify cluster mass assembly history, by probing the accretion phase, and the birth of an object according to its mass growth. 
First, the accretion rate is the ratio between the halo mass at $z=0$ and the mass of its main progenitor at a given $z$ \citep[according the definition of][]{Diemer2013b}:
        \begin{equation}
            \Gamma \equiv \frac{\Delta log(M_{200\text{m}})}{\Delta log(a)} .
            \label{eq:MAR}
        \end{equation}
This parameter allows us to quantify the accretion phase of a given halo between two time steps, chosen to be $z=0$ and $0.5$ \citep[which corresponds to the expected relaxation timescales of halos, according to][]{Power2012,Diemer2014,More2015}.
Secondly, the formation redshift is the time at which the mass of the main progenitor halo is equal to half its mass at the present time: 
    \begin{equation}
    \frac{M_{200}(z_\text{form})}{M_{200}(z=0)} = 1/2 \,,
     \label{eq:zform}
    \end{equation}
following the definition of \cite{Cole1996}.  \\

\textbf{Structural properties.}
The structural properties of clusters have been also estimated based on three different parameters: the center offset $R_{off}$, the subhalo fraction $f_\text{sub}$ and the halo ellipticity $\epsilon$, computed on the 3-D matter distribution inside the virial radius $R_\text{vir}$ of each cluster.
First, the center of mass offset is computed as the distance between the center of mass $r_\text{cm}$ and the density peak $r_\text{c}$ normalised by the virial radius, such as 
\begin{equation}
R_{off} = \vert r_\text{cm} - r_\text{c} \vert / R_\text{vir}.
    \end{equation}
Secondly, the subhalo mass fraction represents the amount of mass contained in sub-clumps hosted inside a halo.
It is defined as the ratio between the sum of all subhalo masses (without taking into account the main subhalo) and the total halo mass $M_\text{tot}$, such as 
    \begin{equation}
    f_\text{sub} = \sum M_\text{sub} / M_\text{tot}.
    \end{equation}
Thirdly, the shape of cluster halo is quantified by measuring the ellipticity of DM distribution in two (and three) dimensions. Following \cite{Suto2016}, the fitted ellipsoid on matter distribution is found by computing the eigenvalues of the mass tensor of all DM particles, and by fixing the total mass enclosed in the ellipsoid equals to $M_{200}$. The two (and three) dimensional ellipticity are
    \begin{equation}
    \epsilon_{2D} = \frac{c - a}{2(a + c)} \,, \\ 
    \epsilon_{3D} = \frac{c - a}{2(a + b + c)} ,
    \end{equation}
with $a$ ($\leq b$) $\leq c$ the major, (intermediate,) and minor axis vectors of the ellipsoid \citep{Jing2002}. 
Fig. \ref{fig:FIG_beauty} illustrates the ellipse modeled following this procedure in the 2-D DM distribution of a given cluster. \\

\textbf{Dynamical state.}
In order to investigate the dynamical state of clusters, the so-called relaxedness parameter $\chi_\text{DS}$ has been computed following the definition of \cite{Haggar2020},
\begin{equation}
    \chi_\text{DS} = \sqrt{ \frac{3}{ \left(\frac{\Delta_\text{r} }{0.07} \right)^2 + \left(\frac{f_\text{sub} }{0.1} \right)^2 + \left(\frac{\eta -1}{0.15} \right)^2 }} .
    \label{eq:DS}
\end{equation}
This equation is a quadratic average of three dynamical and structural proxies, with $\eta$ the ratio between the kinetic energy and the gravitational potential energy.
Groups and clusters with $\chi_{DS} \geq 1$ are supposed to be dynamically relaxed, whereas dynamically perturbed systems have a relaxedness value such as $\chi_{DS} < 1$ \citep[see e.g.][]{Kuchner2020}.
Notice that recently, \cite{Zhang2021_300_DS} have extended the above-mentioned relation to a new threshold-free function to classify cluster dynamical states.\\

\textbf{Connectivity.}
The connectivity, $K$, defines the number of cosmic filaments that connect to clusters. 
In practice, it is computed by counting the number of filaments  intersecting a sphere of $1.5\, R_{200}$ radius around the each cluster centers \citep[similarly to][]{Darragh2019,Sarron2019,Kraljic2020}. In the present study, the filamentary pattern in the whole simulation box is detected based on a cosmic-web skeleton constructed from the Graph model of the algorithm T-ReX \citep{Bonnaire2020,Bonnaire2021} and applied on the 3-D subhalos distribution of the simulation \citep[as explained in][]{Gouin2021}.

\section{Radial gas distribution in cluster environments \label{SEC:GAS}}

We introduce in this section key features of the gas properties as a function of the radial distance from the cluster center.
We consider all the gas cells contained around our $415$ halo samples up to $5 \times R_{200}$ (labeled as \textsc{PartType0} in IllustrisTNG). We focus here on two thermodynamical properties: the temperature $T$ \citep[computed under the assumption of perfect monoatomic gas, as in][]{Galarraga2021} and the $n_{H}$ hydrogen number density which is a direct tracer of the total gas density (directly pre-computed in IllustrisTNG).
We refer to, \cite{Martizzi2019} for an accurate description of gas properties in the different cosmic web environments (voids, walls, filaments, and nodes), and \cite{Galarraga2021} for a complete study of gas thermodynamics inside cosmic filaments; two studies of cosmic gas based on the IllustrisTNG simulation. 
In our case, we focus on the particular case of transition from infalling gas along filaments to the captured gas inside cluster gravitational potential wells.

\begin{figure*}
    \centering
    \includegraphics[width=0.98\textwidth]{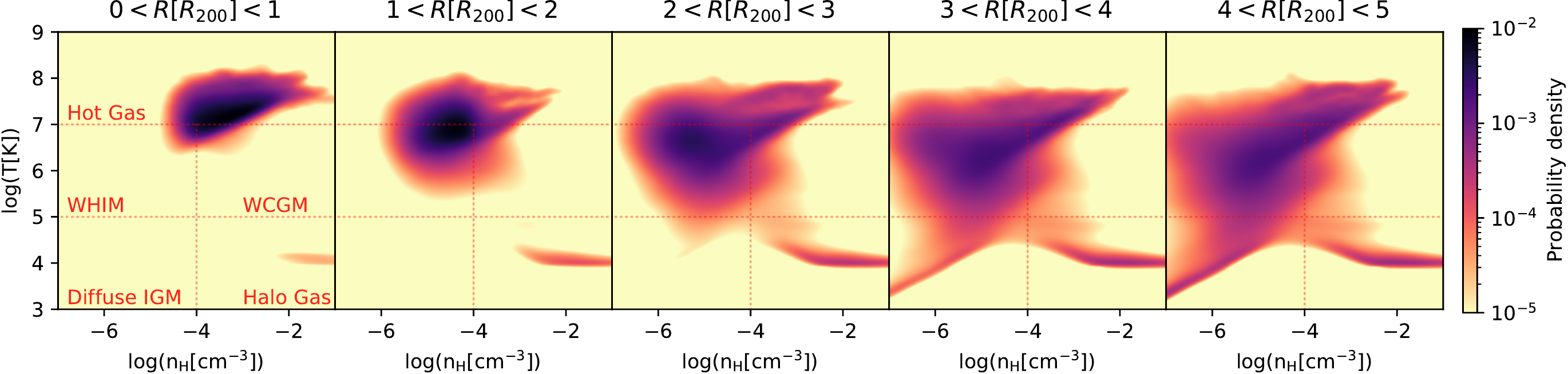}
     \caption{Stacked temperature-density diagrams for all gas cells around galaxy clusters and groups in IllustrisTNG, considering different radial apertures from cluster central regions $R[R_{200}]<1$ up to $4<R[R_{200}]<5$ \label{fig:FIG_T_nH}}
\end{figure*}

\begin{figure}
    \centering
    \includegraphics[width=0.48\textwidth]{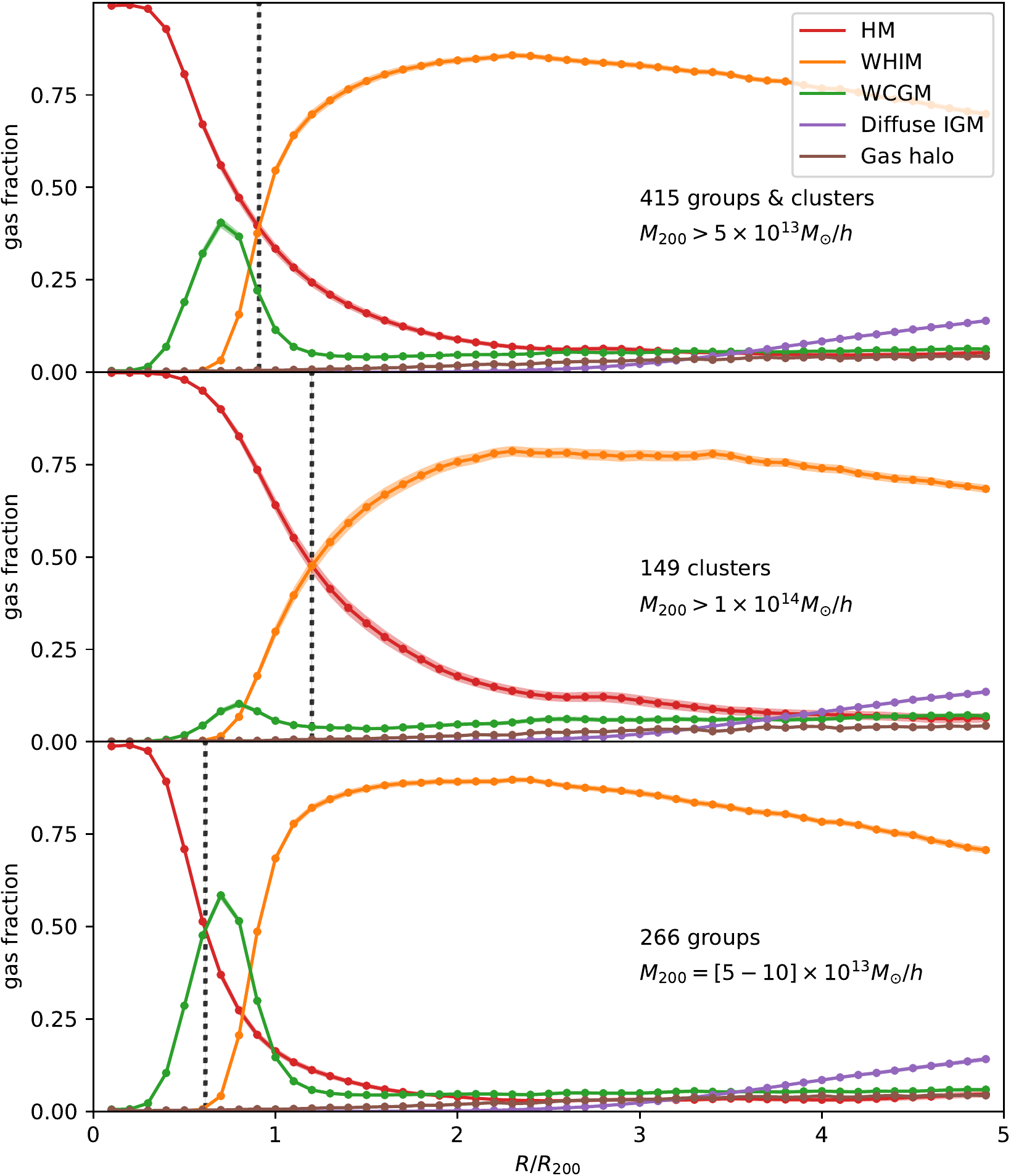}
     \caption{Mean gas mass fraction profiles (see Eq. \ref{EQ:eq_gas_phase}) of the five different gas phases around all halos in our sample (top), only galaxy clusters (middle) and only galaxy groups (bottom). 
     The radial boundary at which the hot gas phase is no longer dominant is shown by gray vertical lines, and is about $R_{\textrm{HOT}} \sim 0.91, 1.20$ and $0.61$ for the all groups and clusters, only clusters, and only groups respectively. \label{fig:FIG_phase_radial_profil}}
\end{figure}

\subsection{The gas phases}

One commonly used way to characterise gas phases is to probe their distribution in a temperature-density diagram, as it allows to artificially separate the gas in different phases \cite[see e.g.][]{Cen2006}. The temperature and density of gas is commonly separated in five gas phases which are related to different environments and physical processes: the diffuse intergalactic medium (Diffuse IGM), the warm-hot intergalactic medium (WHIM), the warm circumgalactic medium (WCGM), the halo gas and the hot gas \cite[see e.g.][for a detailed description of each phase]{Martizzi2019}.
Notice that changing the baryonic physical models in the simulation, in particular excluding AGN feedback, can affect the distribution of gas in the different phases as recently discussed in  \cite{Christiansen2020} and \cite{Sorini2021} \citep[see also][for the influence of radiative models on cluster simulations]{Sembolini2016}.

In Fig. \ref{fig:FIG_T_nH}, we show the stacked gas distribution in temperature-density diagrams of the $415$ galaxy cluster environments in five bins of cluster-centric distances from 0 to $5 \ R_{200}$.
We have normalised the radial aperture centered on clusters by $R_{200}$ to easily stack gas distribution of different clusters with different masses (without mixing their physical radial scale).  
Figure \ref{fig:FIG_T_nH} illustrates the distribution of gas in the different temperature-density phases as a function of cluster radial distance. Inside clusters ($R<1R_{200}$), gas is mainly in the form of hot plasma at high temperature, $T>10^7 K$. Increasing the distance from the cluster centers, from $1$ to $3 \ R_{200}$, we can see that most of the gas is at lower temperature (in the range $10^{5} <T [K]< 10^{7}$) and at lower density (in the range $n_{H}< 10^{4} cm^{-3}$). The gas is transiting from hot dense plasma to diffuse warm gas, in the so-called phase: warm-hot intergalactic medium.
 At larger distances from cluster centers ($>3 R_{200}$), the gas in temperature-density diagrams appears distributed in the different phases: cold diffuse (IGM), cold dense (halo gas), warm diffuse (WHIM), warm dense (WCGM) and hot gas.
 This temperature-density distribution is quite similar to the expected distribution of overall cosmic gas in the universe at $z=0$ \citep[see figure 2 of][which considered all gas cells in the simulation box]{Galarraga2021}. Therefore, it suggests that beyond radial distances larger than $3 \ R_{200}$ the influence of cluster environments is no more significant. An actual quantification based on mass fraction profile, and different cluster mass bins, will be discussed below. \\
 Moreover, according to \cite{Artale2022}, who have probed the large-scale distribution of ionized metals in IllustrisTNG, we can relate these gas density–temperature diagrams in Fig. \ref{fig:FIG_T_nH} to their metal abundance as depending of the cluster-centric distance, and in the context of comparison with UV/optical wavelengths observations. Recently, \cite{Artale2022} found that Mg II, C II, and Si IV are efficient tracers of the halo gas in dense environment, whereas Ne VIII, N V, O VI, C IV are better tracers of warm/hot and low-density gas (WHIM) inside filamentary structure at large scales (and z=0). In particular, the small fraction of Halo Gas phase inside clusters (with $T<10^5 \ K$ and $n_{H}>10^{-4} \ cm^{-3}$) is supposed to be condensed gas inside halos in high-density and low-temperature star-forming regions, as recently observed via Mg II absorption lines by \cite{Lee2021_MGII,Anand2022,Mishra2022}. For example, \cite{Lee2021_MGII} have shown that Mg II absorbers are more abundant inside clusters than outside ($> 2 R_{200}$).

We now probe the detail radial profile of the different gas phases as a function of cluster-centric distance. 
Similarly to \cite{Galarraga2021}, we define the mass fraction of a given gas phase such as:
\begin{equation}
    \psi^{gas}_{i} (r)= \frac{\rho_{i}^{gas}(r)}{\rho_{TOT}^{gas}(r)} \,,
    \label{EQ:eq_gas_phase}
\end{equation}
with $r$ is the 3-D radial distance to the cluster center, $\rho_{i}^{gas}(r)$ the radial density of the $i$th gas phase, and $\rho_{TOT}^{gas}(r)$ the radial density of the total gas.
The radial density of gas is computed by summing the mass of gas cells contained in spherical shells from a radius $r_{k-1}$ to $r_{k}$, following the above equation: 
\begin{equation}
    \rho^{gas}(r_k) = \frac{\sum_{j}^{N_k} m_{j}}{4/3 \pi (r_{k}^3 - r_{k-1}^3)} \,,
    \label{EQ:eq_gas_density}
\end{equation}
with $N_k$ is the number of gas cells $j$ contained in a spherical shell with radius from $r_{k-1}$ to $r_{k}$ centered on the halo. 

Following Eq. \ref{EQ:eq_gas_density} and \ref{EQ:eq_gas_phase}, we have computed the mass fraction of the five gas phases around each halo of our cluster sample.
The mean radial profile of gas phases is presented in Fig. \ref{fig:FIG_phase_radial_profil}, by averaging over our halo sample, only considering galaxy clusters ($M_{200}>10^{14} M_{\odot}/h$) and only considering galaxy groups ($M_{200}=[5-10] \times 10^{13} M_{\odot}/h$) respectively in top, middle and bottom panels.
Figure \ref{fig:FIG_phase_radial_profil} shows that the hot gas phase is strongly dominating the interior of all halos (top panel). As expected the gas must be strongly heated inside the deep gravitational potential wells of clusters, and thus appears in the form of a hot plasma. Beyond the cluster region, the fraction of hot gas decreases, such that the gas becomes warmer and less dense: WHIM gas phase. The warm diffuse gas starts to dominate at distances larger than $\sim 0.9 R_{200}$ on average. 

However, we can see that the transition radius from hot to WHIM gas phase strongly differs for galaxy clusters and galaxy groups (middle and bottom panels).
To illustrate this, we highlight the radius for which hot gas phase is no longer dominant in gray vertical lines in Fig. \ref{fig:FIG_phase_radial_profil}.
The hot plasma extends up to $1.2 R_{200}$ inside galaxy clusters, whereas it is only dominant up to $\sim 0.6 R_{200}$ for galaxy groups. In addition, the hot gas mass fraction inside $1 \ R_{200}$ is about 93\% for galaxy clusters, whereas it is only 68\% for the galaxy groups. While hot plasma is largely extended and represents almost all of the gas inside galaxy clusters; gas in galaxy groups is a mixture of hot and warm dense gas.

The bottom panel of Fig. \ref{fig:FIG_phase_radial_profil} shows that the transition from hot to WHIM gas phase is different in galaxy groups because it implies a third gas phase: the warm circum-galactic medium (WCGM) which has a similar temperature than the WHIM but is denser (see Fig. \ref{fig:FIG_T_nH}).
The shallower gravitational potential of galaxy groups is not deep enough to heat the gas up to $10^{7} K$ beyond the core region ($>0.6 \times R_{200}$). At larger distance, the gas temperature decreases first, with the gas transiting to the warm dense phase (WCGM), and then density decreases at $0.8 R_{200}$, transforming the gas in diffuse and warm phase (WHIM). According to \cite{Martizzi2019}, WCGM gas phase must be created by shock heating and feedback of massive galaxies, and is located mostly at the vicinity of massive galaxies and inside galaxy groups. 
We confirm here that, WCGM phase is one of the dominant phases inside galaxy groups accounting for 23\% of the mass within $R<R_{200}$, with a peak contribution at around $0.7 R_{200}$.

Outside groups and clusters, typically beyond distance $\gtrsim 1 \ R_{200}$, the WHIM gas phase largely dominates, and tends to smoothly peak at around $2 \ R_{200}$. 
These radial distances are typically the infalling regions around clusters where gas is expected to be located inside the cosmic filamentary pattern connected to clusters \citep[see e.g.][for an observational evidence]{Eckert2015}.
In this region, the fraction of hot gas remains still non-negligible, with about 13\% of the total mass of gas from $1$ to $3$ $R_{200}$. In fact, this small fraction of hot gas is expected to be in the form of small massive clumps located in cosmic filaments \citep{Zhuravleva2013,Angelinelli2021}.
Far from the group and cluster centers ($>3 \ R_{200}$), the warm diffuse gas remains the dominated phase. 
Indeed, WHIM gas is supposed to be the dominant phase in the Universe with a mass fraction of around 46\% according to \cite{Martizzi2019}. 
Notice that in our case, at distances of $5\times R_{200}$, the mass fraction of WHIM is quite high with a value of 70\%. This is coherent with the findings of \cite{Galarraga2021}, who have shown that WHIM gas phase slowly decreases up to $20 Mpc$ away from the spine of denser filaments.
We can thus expect similar behaviour for radial WHIM fraction profile around clusters.

We finally notice that, far from cluster centers at around $2.5 \ R_{200}$, the WHIM gas phase starts to decrease in favor of diffuse IGM. It is related to the presence of cold diffuse gas in galaxies. The increase of diffuse IGM at around $2 R_{200}$ from cluster centers, is in agreement with results from \cite{Mostoghiu2021}, who show that cold gas in infalling galaxies is completely depleted at $1.7 R_{200}$ from cluster centers \citep[see also][on the impact of filaments and cluster environments on the depletion of cold gas in infalling galaxies]{Arthur2019,Singh2020,Zhu2021,Song2021}.

\begin{figure*}
    \centering
    \includegraphics[width=0.98\textwidth]{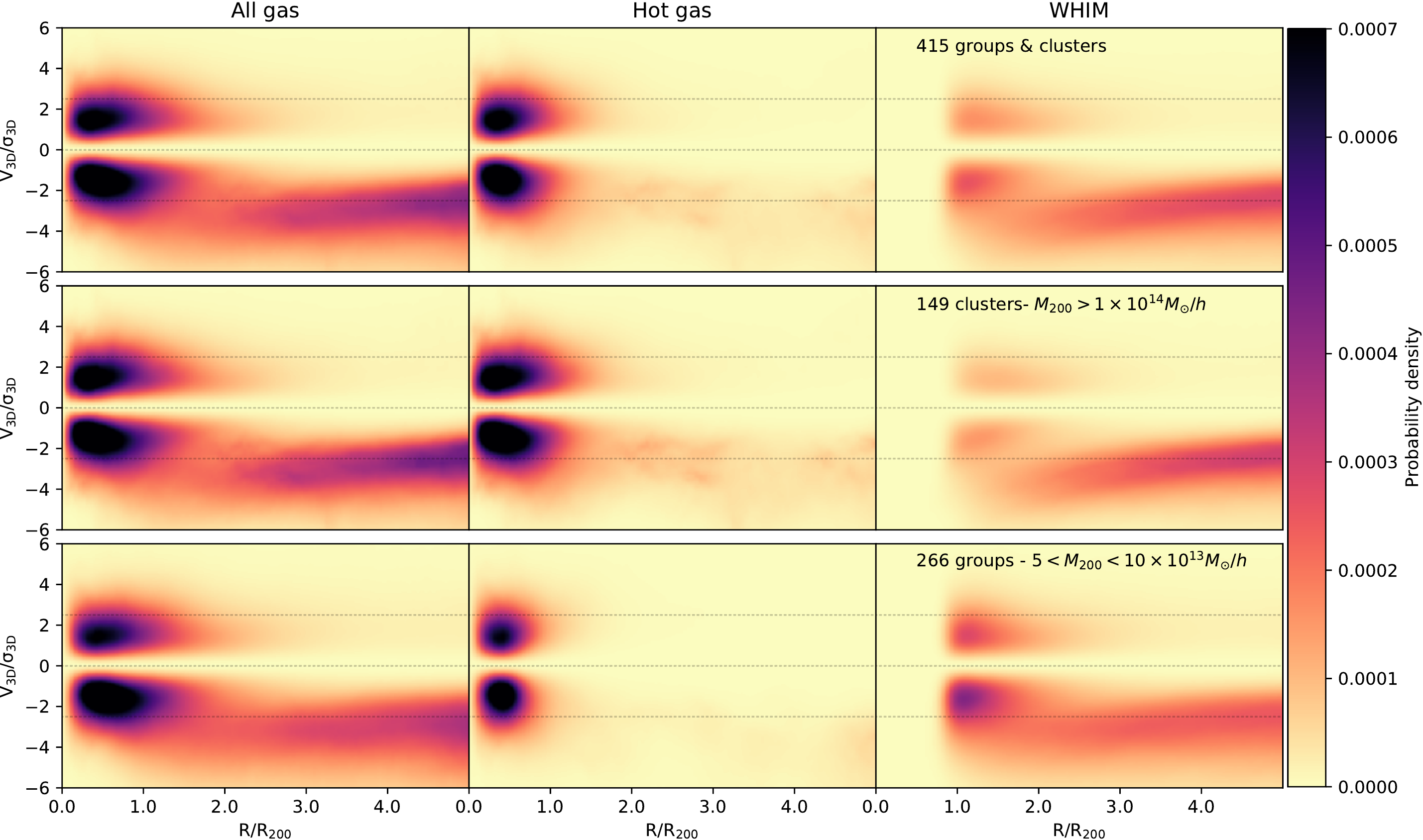}
     \caption{Stacked phase-space diagrams for all gas cells (left panels), only hot gas (middle panels), and warm-hot inter-galactic medium (right panels). All groups and clusters are considered in top panels, only galaxy clusters (with $M_{200}>10^{14} M_{\odot}/h$) in the middle panels, and only galaxy groups (with $M_{200}=[5-10] \times 10^{13} M_{\odot}/h$) in bottom panels.  \label{fig:FIG_V_R}}
\end{figure*}

\subsection{The gas dynamics}

Beyond the temperature-density gas phases, it is also of prime importance to probe gas motions in order to investigate their infall from large-scale environment to clusters.
The dynamics of the gas is explored in phase-space coordinates (velocity, position) following the definitions presented by \cite{Oman2013}.
Considering the 3-D vector position \textbf{r} and 3-D vector velocity \textbf{v} of each gas cell, we can identify all gas cells by their 6-D coordinates $(r_x,r_y,r_z,v_x,v_y,v_z)$.
At the same time, each cluster is characterised by its central position $\boldsymbol{r_c}$, and its 3-D velocity $\boldsymbol{v_c}$ (the halo velocity is computed as the sum of the mass-weighted velocities of all particles/cells in the halo, pre-computed by the IllustrisTNG collaboration).  
Considering these phase-space coordinates, we can define the radial velocity of each gas cell, relative to their associated host cluster environment, such as:
\begin{equation}
    v_{3D} = \textrm{sgn}((\boldsymbol{r}-\boldsymbol{r_c})\boldsymbol{.}(\boldsymbol{v}-\boldsymbol{v_c})) \vert \boldsymbol{v}-\boldsymbol{v_c} \vert  \,.
    \label{EQ_v3d}
\end{equation}
Note that the sign of $v_{3D}$ allows to distinguish between the infalling gas ($v_{3D}<0$) and the outgoing gas ($v_{3D}>0$) around a given cluster. 
In order to stack gas radial velocities for all cluster environments, we normalize them by the overall velocity dispersion $\sigma_{3D}$, which is the root-mean-square of the radial gas velocity $v_{3D}$ within $R_{200}$ of each cluster.

The velocity-position diagram is commonly used to probe the infall of galaxies into clusters \citep{Dacunha2021}. We can consider typically three different dynamical regimes in the velocity-position diagram: (i) infalling material with $v_{3D}<0$ and $R>R_{200}$, (ii) backsplash material with $v_{3D}>0$ and $R>R_{200}$, and (iii) virialized material inside clusters with $R<R_{200}$ characterised by shell crossing caustics in velocity-position diagrams. In this frame, galaxies start their infall far from cluster center and increase their velocity as they approach the cluster. When they are close to the cluster, they first infall and then move away from the cluster (to form the backsplash population), and then they transit between infall and backsplash, until they are captured by the gravitational potential wells of cluster, forming the virialised population ($R<R_{200}$) \citep[see figure 2 of][for a schematic view of phase-space plane]{Arthur2019}.

In Fig. \ref{fig:FIG_V_R}, we present phase-space diagram for gas in cluster environments. They are obtained by stacking all gas cells (left panels), only hot gas (middle panels), and only WHIM gas (in right panels) for the $415$ halos of our sample (top panels), only the $149$ galaxy clusters with $M_{200}>10^{14} M_{\odot}/h$ (middle panels), and only the galaxy groups with $M_{200}=[5-10] \times 10^{13} M_{\odot}/h$ (bottom panels).
In the top left panel, the stacked phase-space diagram of all gas cells in all cluster environments shows the three dynamical regimes: virialised gas inside clusters, infalling gas with high infalling velocity, and splashback gas component with positive velocity outside clusters \citep[identically to][for galaxy distribution]{Mostoghiu2019,Dacunha2021}.
In more details, we see that the virialised gas inside clusters ($R<R_{200}$) is the hot gas phase, as shown in top middle panel. This hot plasma is virialised, i.e. there is an equal balance between inflow and outflow motions with low velocity values $-2.5 \sigma<v_{HM}<2.5 \sigma$.
Focusing on the WHIM gas phase in top right panel, we can distinguish two distinct dynamical behaviors. 
At cluster peripheries, from $\sim 1$ to $\sim 2 \ R_{200}$, the WHIM gas is a combination of slowly infalling and backsplash gas ($-2 \sigma<v_{WHIM}<2 \sigma$). In contrast to hot gas, the WHIM inflow and outflow motions are not balanced and infalling gas is significantly dominant. 
This means that this WHIM gas is accumulating at cluster peripheries, by slowly infalling inside clusters and with a minor fraction which is ejected outside clusters.
Beyond the cluster peripheries, at distances from $1.5$ to $5 R_{200}$, the WHIM gas is also rapidly infalling on clusters with high velocity values, typically $-5 \sigma<v_{WHIM}<-2 \sigma<v_{3D}$. 
According to \cite{Rost2021}, gas is entering into clusters with turbulent motions. This might explain that WHIM gas phase is accumulated and ejected at cluster peripheries ($\sim 1R_{200}$). Moreover, \cite{Rost2021} have also shown that gas is preferentially infalling from filaments. Therefore, one can suppose that fast infalling WHIM ($R>R_{200}$ and $v_{WHIM}<-2 \sigma$) must be inside cosmic filaments whereas the WHIM backsplash material ($\sim R_{200}$ and $v_{WHIM}>0$) should leave the cluster centre outside filaments.

Regarding the halo mass dependency, we can compare middle and bottom panels showing clusters and groups of galaxies, respectively. 
Focusing on all gas cells, there are not significant differences between clusters and groups. 
However considering hot and WHIM gas phases separately, we can see that galaxy groups and clusters have different dynamical behaviors. 
The hot gas is more spatially extended inside galaxy clusters compared to groups, as already found in the radial gas phase profile (Fig. \ref{fig:FIG_phase_radial_profil}).
Also, we can notice that only galaxy clusters present residual hot gas distribution beyond $R_{200}$. This small fraction of infalling hot gas must be associated with dense clumps of matter inside cosmic filaments as proposed by \cite{Angelinelli2021}. 
Regarding the WHIM velocity-position diagram, we can see that the warm gas dynamics is different for groups and clusters. 
Whereas the WHIM gas phase is mostly in the form of fast infalling gas (high velocity and low velocity dispersion) outside galaxy clusters, WHIM gas around groups is strongly accumulating at their outskirts (with large scatter in velocity, and backsplashing motion from $1$ to $2 \ R_{200}$).
This suggests that WHIM gas phase is mostly inside filaments at cluster peripheries, whereas WHIM gas around groups is infalling and ejected at cluster peripheries. \\

\textbf{Radial gas properties summary}

We conclude that the hot gas is the dominant phase inside cluster halos ($R<R_{200}$), mostly governed by low velocity dispersion of well-balanced inflow and outflow motions ($\vert v_{HM} \vert \lesssim 2.5 \sigma$). 
In contrast, outside groups and clusters ($R> R_{200}$) the gas is mainly in the form of warm diffuse gas. The WHIM gas phase outside halos can be separated in two distinct dynamical regimes, accumulating gas slowly infalling from $\sim 1$ to $\sim 2 \ R_{200}$, and fast infalling gas at distances larger than $1.5 \ R_{200}$.
Cluster peripheries are thus crucial places to probe the transition between hot and warm gas, and to probe the complex dynamics of infalling WHIM gas. In fact, it is in these regions that the gas is expected to flow from filaments into clusters, with turbulent motions \citep{Rost2021,Valles2021}, and where accretion shocks might arise \citep{Shi2020,Zhang2020,Zhang2021,Zhu2021_shock}.
Following these findings on the radial gas properties, we focus in the rest of the paper our exploration of the azimuthal gas distribution in cluster environment concentrating on two main gas phases in two different radial apertures: hot medium up to $1 \ R_{200}$ and WHIM from $1$ to $2 R_{200}$.

\section{Azimuthal distribution as a proxy of structural properties of clusters \label{SEC:MULTIPOLE}}
\begin{figure*}
    \centering
    \includegraphics[width=0.98\textwidth]{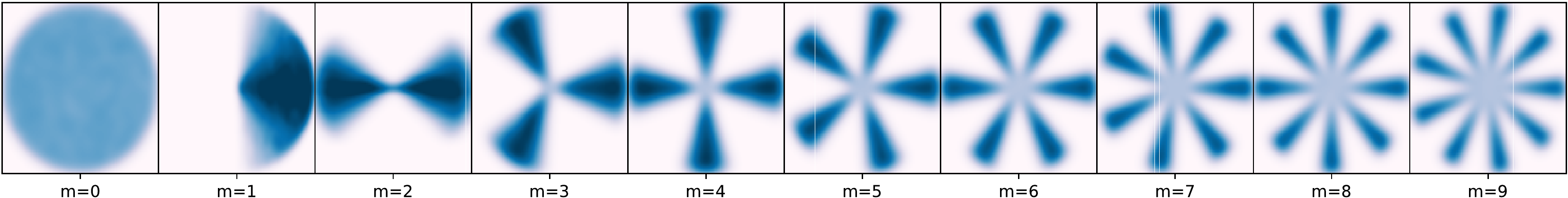}
    \caption{Illustration of the different azimuthal symmetries as a function of harmonic orders $m$. \label{fig:Annex_0}} 
\end{figure*}

\begin{figure*}
    \centering
    \includegraphics[width=0.98\textwidth]{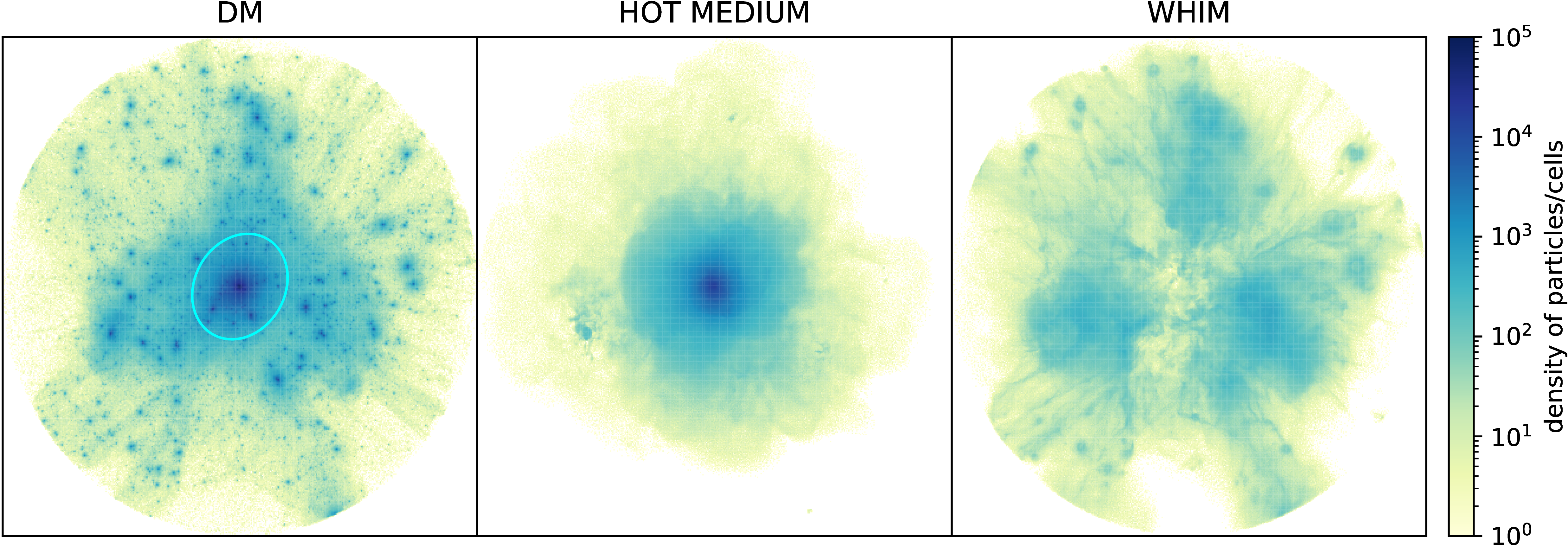}
     \caption{Illustration of the dark matter (left), hot gas (middle) and WHIM (right) 2-D projected distribution up to $5 R_{200}$ of one simulated galaxy cluster. The cyan ellipse on dark matter traces the ellipticity of the DM halo ($\epsilon^{2D}_{DM}$) as discussed in \ref{SEC:CLUSTER}. \label{fig:FIG_beauty}
     }
\end{figure*}

In this section, we define the aperture multipole formalism, and use this technique on gas azimuthal distribution to statistically highlight angular-dependent features in comparison to the \textit{"reference"} DM distribution. This technique focuses on 2-D spatial distribution, and has been already applied on on weak lensing maps \citep{Dietrich2005} and on projected photometric galaxy distribution \citep{Gouin2020}.
In this study, the azimuthal symmetries of gas and DM are explored as a function of cluster structural properties (defined in Sect. \ref{SEC:CLUSTER}), to probe if they are good tracers of the structural features of cluster halos.

In this section and the next one, a general colour and style code is used in the plots such as: 
(i) Hot gas, WHIM, and dark matter are respectively in red, orange and blue/black colors, 
(ii) the mean profiles are shown by solid lines, and the errorbars are the errors on the mean computed by bootstrap re-sampling, 
(iii) the number of objects used to compute the average in each bin (of x-axis) is written on the top of the figures in gray,
(iv) the Spearman's Rank correlation coefficient $\rho_\text{sp}(X,\beta_m)$ is written on the figure, with $X$ the cluster property and $\beta_m$ the multipolar ratio at the multipole order $m$ as defined below. The \textit{p}-value of the correlation coefficient remains lower than $10^{-3}$ for each plots. 

\subsection{Formalism of multipole moments $Q_m$}

The aperture multipole moments of 2-D density fields was first introduced by \cite{Schneider1997} for weak lensing map application. It consists in a multipole decomposition of a surface density in harmonics modes $m$, integrated over a given radial aperture, such as:
\begin{equation}
     Q_m (\Delta R) = \int^{R_{max}}_{R_{min}} \int_0^{2\pi}   R \ dR  \ d \phi \ e^{im\phi} \ \Sigma (R,\phi) \, ,
     \label{eq:multipole_qm}
\end{equation}
where $\Sigma(R,\phi)$ is the 2-D matter distribution in polar coordinates centered on the cluster center. The 2-D radial aperture $\Delta R=(R_{max},R_{min})$ is also centered on the cluster center with $R_{min}$ and $R_{max}$ the radii delimitating the circular shell. For illustration, we show the different azimuthal symmetries quantified by the multipole moments $Q_m$ in Fig. \ref{fig:Annex_0} with monopole ($m=0$), dipole ($m=1$), quadrupole ($m=2$), etc.
Using the multipolar expansion of 2-D matter distribution around galaxy clusters, this technique succeeded in highlighting both the elliptical shape of clusters \citep[see e.g.][]{Clampitt2016B,Shin2018}, and filamentary patterns at cluster outskirts \citep{Dietrich2005,Mead2010,Gouin2017,Codis2017} from 2-D projected DM distribution. 

In order to assess the relative weight of one azimuthal symmetry traced by the order $m$ relatively to the circular one, we define the multipolar ratio $\beta_m$ such as
\begin{equation}
    \beta_m =\frac{ \vert Q_m \vert }{ \vert Q_0 \vert} \,,
    \label{eq:multipole_beta}
\end{equation}
with $\vert Q_m \vert$ the modulus of the aperture multipole moment at the order $m$.
By decomposing surface mass density in harmonic expansion terms, \cite{Schneider1991} have shown that $\frac{\vert Q_m \vert}{\vert Q_0 \vert} \longrightarrow 1/2$ for a matter distribution $\Sigma(\theta) \propto \cos(2\theta m)$. Therefore, the value of the multipolar ratio $\beta_m$ varies between $0$ and $1/2$, such that $\beta_m=0$ for a circular matter distribution and $\beta_m=1/2$ for a matter distribution describing the azimuthal symmetry at the order $m$ \citep[see also][for a similar definition]{Valles2020}.

Here, we aim at probing different aspherical features from the dipolar signature traced by $m=1$, to large harmonic orders (small angular scale patterns) up to $m=9$. Indeed, \cite{Gouin2020} showed that azimuthal matter distribution away from the cluster centers can be described by the sum of multipolar moments from $m=1$ to $m\sim9$, tracing multi-angular scale filamentary patterns. 
We present in appendix Fig. \ref{fig:Annex_1} the evolution of the different multipolar ratio $\beta_m$ as a function of radial distance for $m=1$ to $9$. As expected, the  multipolar ratio increases with the radial distance, similarly to the more commonly used azimuthal scatter technique \citep{Eckert2015}. Indeed, anisotropy of matter distribution is expected to increase with the radial distance from the halo center, as also detailed in \cite{Despali2017} for the ellipticity term.

From the equations \ref{eq:multipole_qm} and \ref{eq:multipole_beta}, we have computed the multipolar ratio $\beta_m$ for $m$ from $1$ to $9$, in both (hot and WHIM) gas and DM distribution for each cluster. 
We have projected the matter distribution enclosed inside a sphere of radius $R_{3D} = 5 R_{200}$ centered on each cluster considering the three different projection axes (along $x$-, $y$- and $z$- axis) of the simulation box.
Therefore, the cluster sample increases from $415$ clusters to $1245$ 2-D projected maps of cluster mass distribution. 
In Fig. \ref{fig:FIG_beauty}, we show an example of the DM, the hot medium and the WHIM 2-D projected distribution around a simulated cluster. 
As discussed in Sect. \ref{SEC:GAS}, the hot gas is mostly located inside the cluster halo (typically $<R_{200}$), whereas WHIM traces the infalling gas outside cluster. In the following, we will thus focus on the hot medium to explore the gas inside clusters, and the WHIM to probe the gas filamentary pattern outside clusters.

\begin{figure*}
    \centering
    \includegraphics[width=0.49\textwidth]{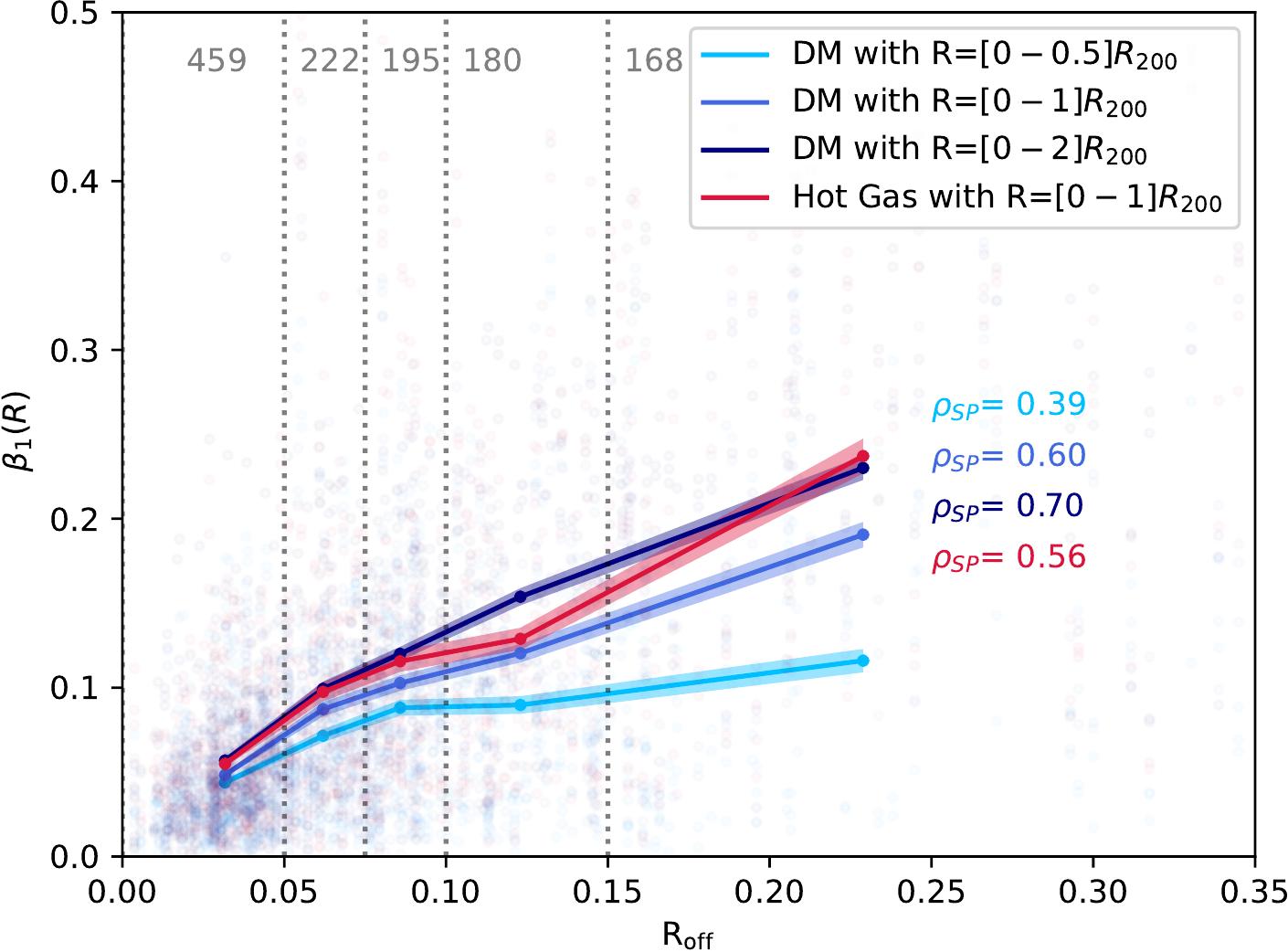} \includegraphics[width=0.48\textwidth]{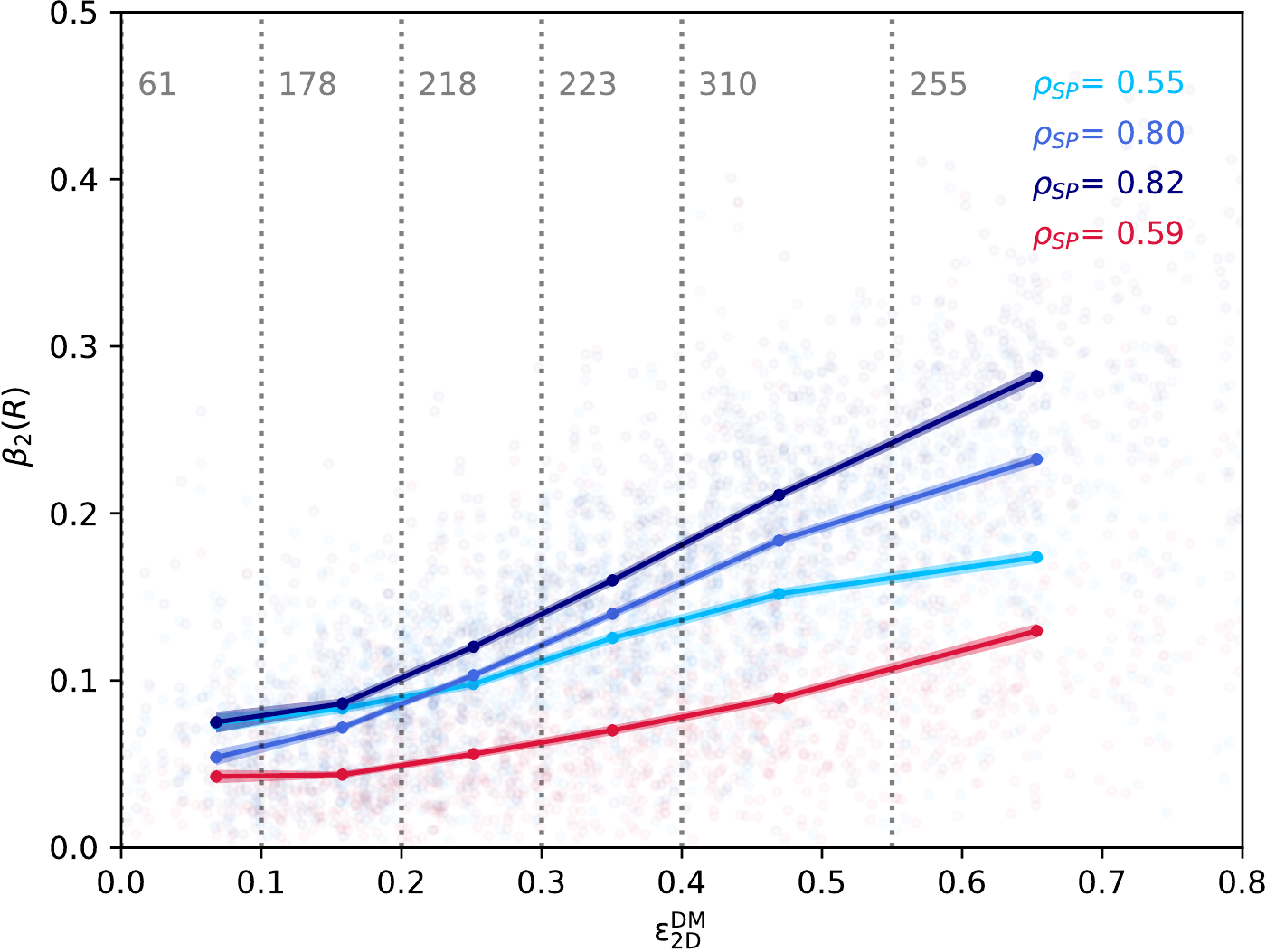} \\
    \includegraphics[width=0.48\textwidth]{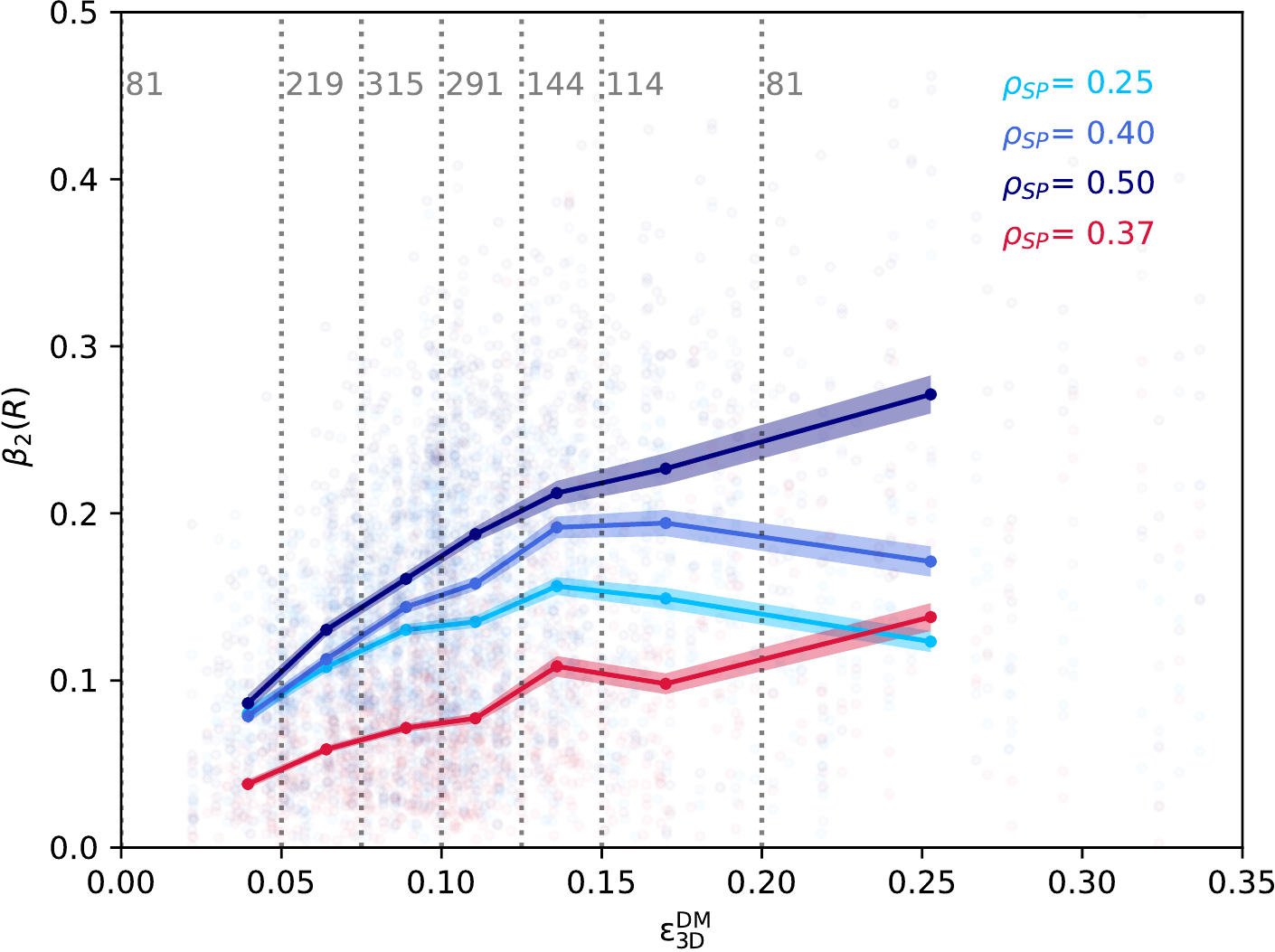} \includegraphics[width=0.48\textwidth]{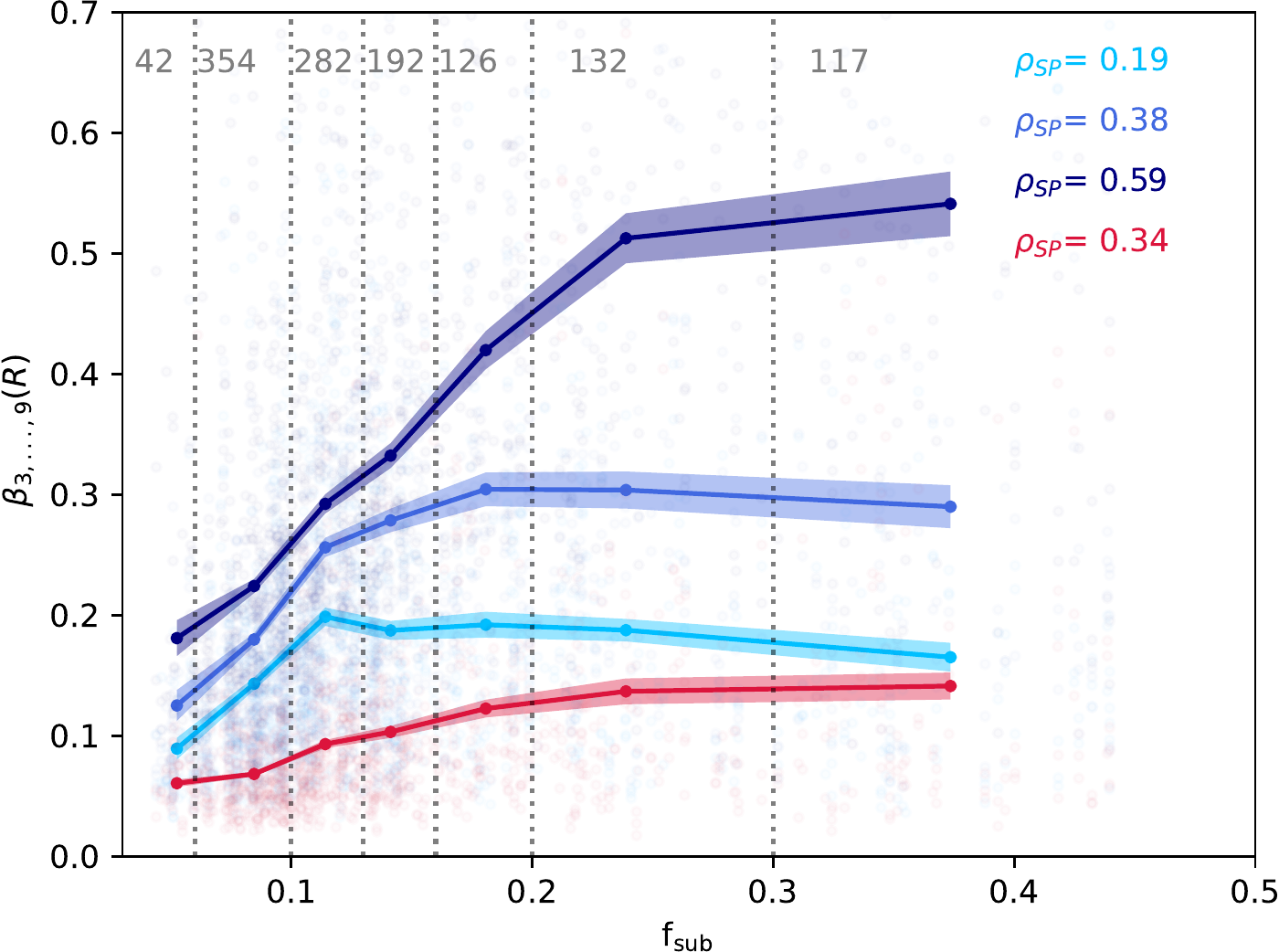}
    \caption{Top left panel: Distribution of $\beta_m$ dipole contribution ($m=1$) as a function of the center of mass offset. 
    Top right panel: Distribution of $\beta_m$ quadrupole contribution ($m=2$) as a function of the 2-D ellipticity of DM.
     Bottom left panel: Distribution of $\beta_m$ quadrupole contribution ($m=2$) as a function of the 3-D ellipticity of DM. 
     Bottom right panel: High azimuthal symmetries $\beta_m$ contribution (summing contributions from $m=3,4,5,6,7,8,9$) as a function of the mass fraction of substructures. \\
     The mean profiles of $\beta$ are shown by solid lines, and the errorbars are the errors on the mean computing by bootstrap re-sampling. 
     The $\beta$ values for the DM distribution with apertures $R<0.5 \times R_{200}$,$R<1 \times R_{200}$ and $R<2 \times R_{200}$, are respectively plotted in light, medium, and dark blue. 
     The $\beta$ values for 2-D hot gas distribution with $R<R_{200}$ is plotted in red.
     On each panel, the number of objects used to compute the average in each bin of x-axis (shown in gray dotted lines) is written on the top of the figures in gray}. 
    \label{fig:FIG_STRUCURAL}
\end{figure*}

\subsection{Impact of cluster structural properties}

Let us now explore the azimuthal symmetries traced by hot gas and DM distribution inside clusters, and see if they are related to the structural properties of cluster halos. 
To do so, we focus here on different angular properties of matter distribution by considering the multipolar ratio $\beta_m$ at different multipolar order $m$. 
Following the description in Sect. \ref{SEC:CLUSTER}, the structural properties we will consider are: the halo ellipticity, $\epsilon$, the center offset, $R_{off}$, and the mass fraction of substructures, $f_{sub}$.

In Fig. \ref{fig:FIG_STRUCURAL}, we show the multipolar ratio $\beta_m$ and its mean, considering different radial apertures $R$, and as a function the structural properties of cluster halo. The multipolar moments and their ratios have been computed for both DM (at different radial apertures) and hot gas (with radial aperture $R<R_{200}$), and are in blue and red, respectively. 

In the top left panel, we show the possible correlation between the center offset and the dipolar symmetry traced by $\beta_1$. We focus on $m=1$, because the dipolar symmetric excess must in principle reflect the mis-centering of mass distribution, whereas the offset center parameter directly quantifies differences between peak center and center of mass. 
We find that the dipolar symmetries of DM and hot gas trace well the center offset of clusters, with a correlation coefficient between $\beta_1$ and $R_{off}$ larger than $\rho_{SP}>0.5$.
Notice that, as expected by taking into account larger mass distribution up to $2 \ R_{200}$ we increase the correlation between dipolar ratio $\beta_1$ and center offset.

The top right panel shows the level of quadrupolar symmetry (with $m=2$) as a function of the 2-D ellipticity of DM halos. 
In agreement with previous studies \citep{Clampitt2016B,Shin2018}, we confirm that the quadrupolar decomposition of the DM density field is strongly correlated with the elliptical shapes of the halo with $\rho_{SP}\sim0.8$.
Notice that increasing the radial aperture above $R_{200}$ does not significantly increase the correlation coefficient, suggesting that the halo shape information is mostly contained inside $R_{200}$. 
We also found a good correlation between the quadrupolar symmetry of hot gas and the DM ellipsoidal shape.
In fact, \cite{Okabe2018} have shown that the gas distribution follows the elliptical shape of DM but tends to be more circular due to the dissipative baryonic processes \citep[see also][]{Velliscig2015}.
In the present study, we find that the hot gas quadrupolar signature is smaller than the DM ($\beta^{DM}_2 > \beta^{hot \ gas}_2$), confirming that the hot gas distribution tends to be more circular than the DM one. This is  also illustrated in Fig. \ref{fig:Annex_ellipticity} in appendix and it confirms relations between both hot gas and DM ellipticity, and their quadrupoles.

In the bottom left panel of Fig. \ref{fig:FIG_STRUCURAL}, we extend our investigation to the 3-D elliptical halo shape. As expected, the correlation between ellipticity and quadrupole remains strong, even if high 3-D ellipticity ($\epsilon_{3D}>0.15$) is degenerate with the 2-D quadrupolar ratio. Indeed due to projection effects, a strongly-elliptical halo shape can produce a low quadrupolar signature. This degeneracy is reduced by considering larger radial apertures, such that for a radial aperture of $2 R_{200}$, the quadrupolar ratio and 3-D ellipticity are strongly correlated with a Spearman coefficient $\rho_{SP} \sim 0.5$.

In the bottom right panel of Fig. \ref{fig:FIG_STRUCURAL}, we finally probe the amount of mass inside sub-halos via the fraction of substructures. 
To quantify this last structural property, the best choice is to consider larger harmonic orders meaning small angular scale decomposition. We directly sum multipolar ratios $\beta_m$ from $m=3$ to $9$ to obtain the overall level of azimuthal symmetries for small angular scales, and compare it to the fraction of substructures. 
Multipolar ratio and $f_{sub}$ tend to correlate well for both hot gas and DM with correlation coefficients around $\rho_{SP}\sim 0.4$. In particular, we can distinguish between small ($f_{sub}<0.1$) and high substructure fractions ($f_{sub}>0.1$) which is an important criterion to separate dynamically relaxed and non-relaxed clusters as proposed by \cite{Cui2017}. Notice that, same as for 3-D ellipticity, by considering large radial apertures ($R<2R_{200}$), we increase significantly the correlation between the fraction of substructures and the level of azimuthal symmetries up to $\rho_{SP}\sim 0.6$. \\

\textbf{Summary of multipole mode decomposition}

As illustrated in Fig. \ref{fig:Annex_0}, the multipole moment at each order $m$ represents different azimuthal symmetries. By probing the multipolar ratio $\beta_m$ (defined in Eq. \ref{eq:multipole_beta}), we have computed the excess of one azimuthal symmetry at the order $m$ relative to the circular symmetry ($m=0$). At each harmonic order $m$, the multipolar ratio highlights a different angular feature. They are by nature powerful tracers of distinct structural properties of cluster halos (as shown in Fig. \ref{fig:FIG_STRUCURAL}). Considering $m=1$, the monopolar ratio reflects the mis-centering of cluster mass distribution, whereas for $m=2$ the quadrupolar excess correlates with the elliptical shape of clusters \citep[as also discussed in][]{Gouin2017,Gouin2020}. Moreover, as we increase the multipole order $m$, the physical size of the angular pattern decreases, and thus, we characterise small scale structures. This explains why multipole modes from $m=3$ to $9$ are a good probe of the substructure fraction.

Even when considering 2-D matter distribution, the correlation with its 3-D structural properties is strong, and can be further improved by integrating matter distribution beyond $R_{200}$. Moreover, the azimuthal hot plasma distribution appears to follow well the azimuthal DM distribution, as shown via its significant correlation with the halo properties. The hot gas plasma distribution remains smoother and more circular than DM with lower values of multipolar ratio $\beta_m$ for almost all orders $m$.

\section{Azimuthal distribution related to cluster physical properties \label{SEC:MAH}}

In this section, we will not attempt to distinguish between the different angular features in the 2-D mass distribution, but we will rather assess whether the overall departure from circular symmetry can be related to cluster physical properties.
We thus focus on a single variable $\beta$ to estimate the amount of azimuthal symmetries in excess compared to the circular one:
\begin{equation}
    \beta=\sum_{m=1}^{N} \beta_m .
    \label{EQ:BETA_1to4}
\end{equation}
This azimuthal symmetric excess, $\beta$, is defined as the sum of all the azimuthal symmetry contributions up to the order $N$. As discussed in Appendix \ref{APPENDIX_BETA}, we have chosen $N=4$ because adding larger orders (from $5$ to $9$) do not alter the conclusions of this section. 
Indeed, each multipole order from $m=5$ to $9$ contributes to less than 10\% to the matter multipolar expansion, and thus, they constitute minor harmonic orders to represent matter distribution (as shown in Fig. \ref{fig:Annex_1}).

Based on the analysis of the radial gas distribution discussed in Sect. \ref{SEC:GAS}, we focus on two radial apertures and two gas phases. 
First, we consider the interior of clusters ($R<R_{200}$) and we investigate the azimuthal symmetric excess $\beta$ of both hot plasma and DM distributions. Secondly, we consider the cluster peripheries ($1<R [R_{200}]<2$) and investigate the filamentary patterns in the WHIM and DM azimuthal distributions via the $\beta$ variable. 
Notice that we have also studied the azimuthal matter distribution for larger radial apertures from $2$ to $5 \ R_{200}$. We found a weak correlation between the cluster physical properties and the azimuthal gas distribution for apertures $2<R [R_{200}]<3$, and the two quantities become uncorrelated beyond radial distances of $3 R_{200}$. 

\begin{figure}
    \centering
    \includegraphics[width=0.48\textwidth]{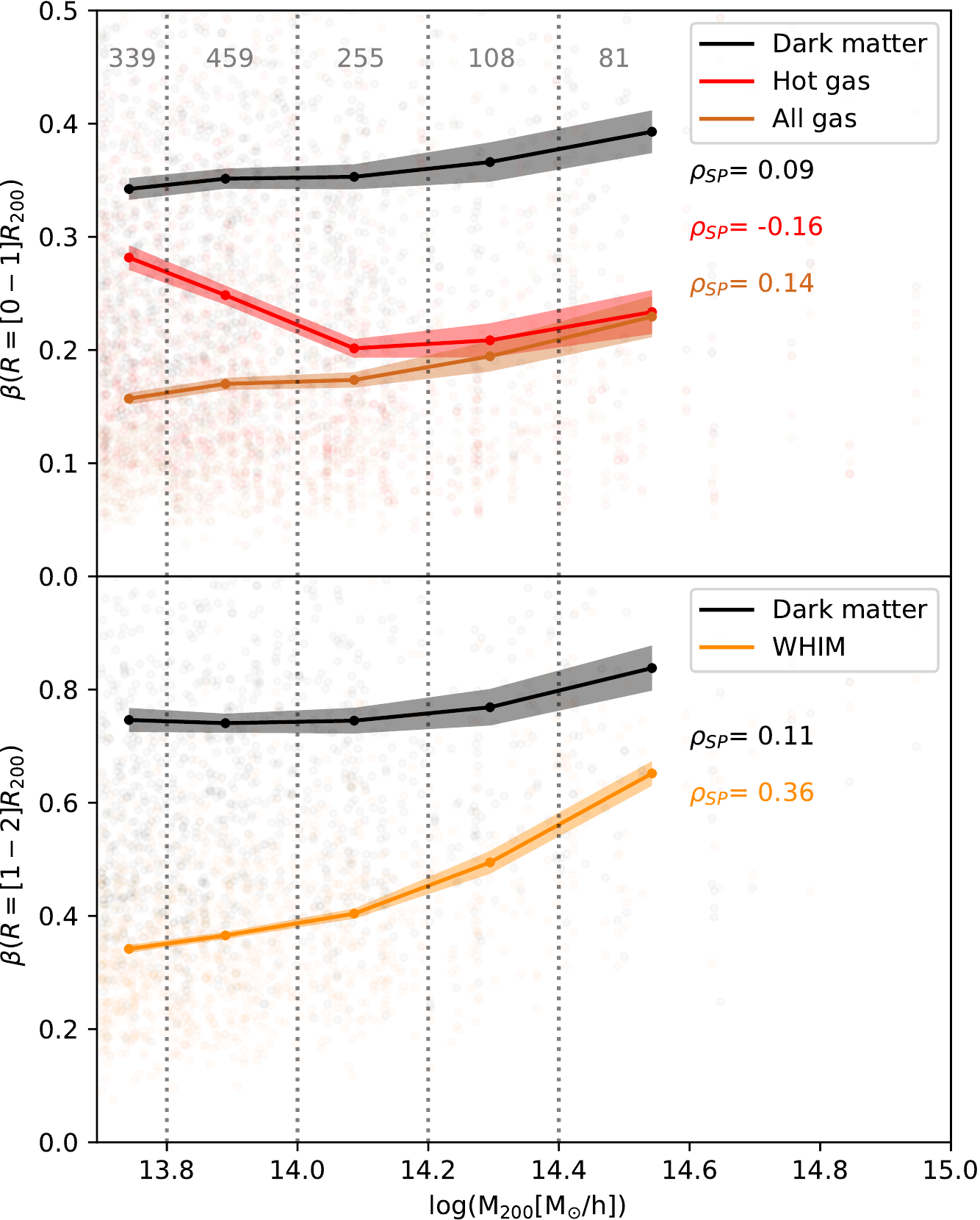}\\
    \caption{Distribution of the azimuthal symmetric excess $\beta$ (as defined in Eq. \ref{EQ:BETA_1to4}) as a function of the halo mass, inside clusters ($R<R_{200}$) in the top panel, and at cluster peripheries in ($1<R [R_{200}]<2$) in the bottom panel.  The mean profiles of $\beta$ and their errors are shown in solid lines. The color of points and lines represent different matter component: dark matter (black), hot gas (red), WHIM (orange), and all gas (light brown). 
    On each panel, the number of objects used to compute the average in each bin of x-axis (shown in gray dotted lines) is written on the top of the figures in gray}. 
    \label{fig:FIG_MASS}
\end{figure}

\begin{figure}
    \centering
    \includegraphics[width=0.45\textwidth]{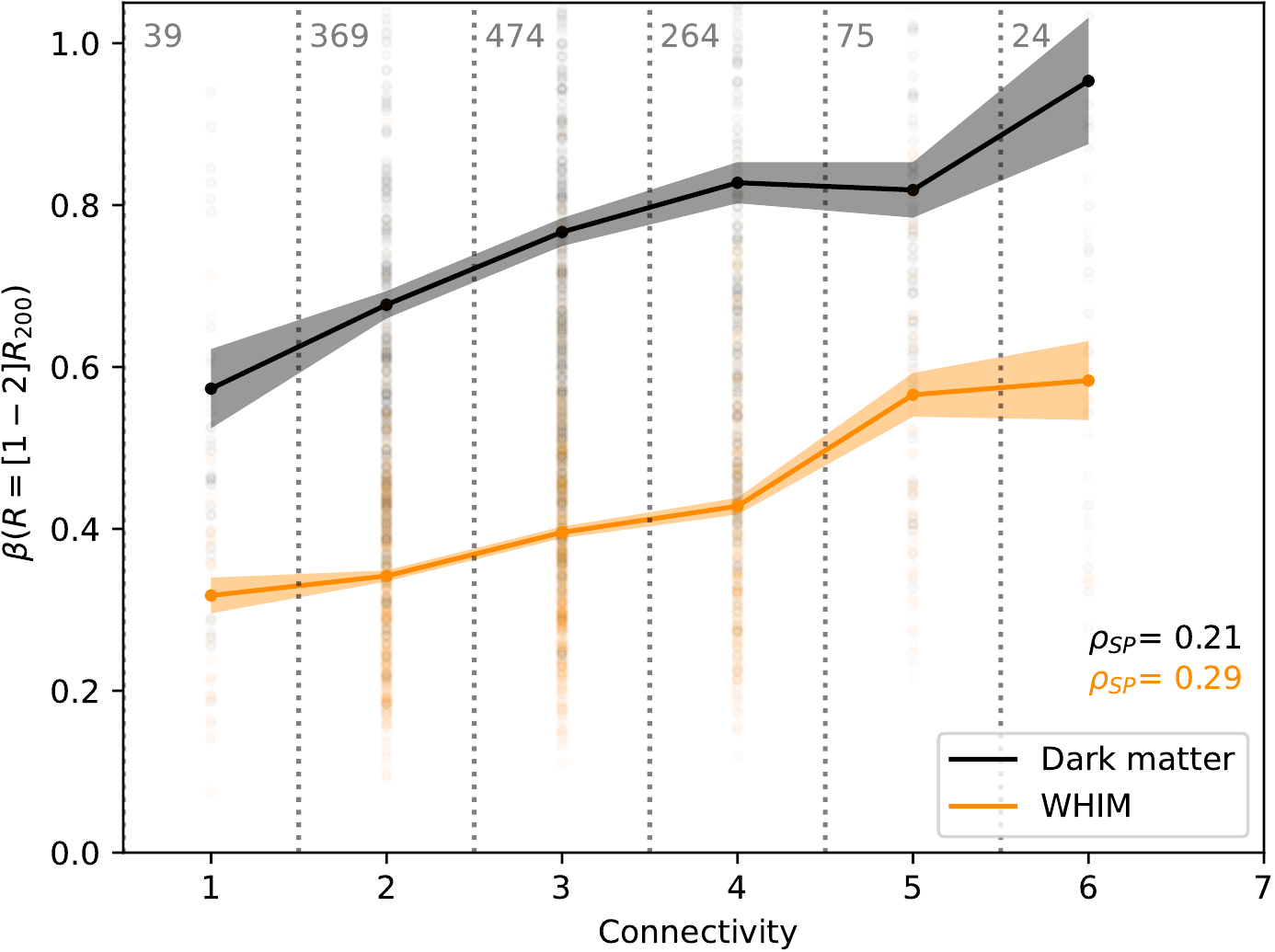}
    \caption{Distribution of the azimuthal symmetric excess $\beta$ (as defined in Eq. \ref{EQ:BETA_1to4}) computed at cluster peripheries in ($1<R [R_{200}]<2$) as a function of the halo connectivity, for DM (black) and WHIM (orange). The mean profiles of $\beta$ and their errors are shown in solid lines. 
    The number of objects used to compute the average in each bin of x-axis (shown in gray dotted lines) is written on the top of the figures in gray}. 
    \label{fig:FIG_Connect}
\end{figure}

\subsection{Mass and connectivity dependency}

 We first discuss the relation between departures from circular symmetry and the cluster mass inside ($R<R_{200}$) and at the cluster outskirts ($1<R[R_{200}]<2$) as shown in top and bottom panels of Fig. \ref{fig:FIG_MASS}, respectively. 

Focusing on the dark matter distribution inside halos ($R<R_{200}$), we note that the azimuthal symmetric excess slowly increases with the cluster mass, on average. In fact, the non-circularity quantified by $\beta$ is strongly dominated by the quadrupole ($m=2$) inside clusters, and traces the elliptical halo shape, as discussed previously in Sect. \ref{SEC:MULTIPOLE} and illustrated in Fig. \ref{APPENDIX_ELLIPSE}. Therefore, the increase of the azimuthal symmetric excess with the halo mass must reflect the increase of the DM halo ellipticity, in agreement with \cite{Despali2014}. Nevertheless, we found that the correlation between the DM azimuthal symmetric excess and halo mass is low, with $\rho_{SP}\sim 0.1$.

Focusing on the hot gas inside clusters, we found that the anisotropy of hot gas distribution tends to decrease with the halo mass. Hot gas in galaxy groups ($M_{200}<10^{14} M_{\odot}/h$) appears more asymmetric than in massive clusters, on average. This might be explained be the fact that, as detailed in Sect. \ref{SEC:GAS}, hot gas is not the dominant component in groups (representing only 68\% of all gas) and is concentrated in the core of galaxy groups (hot gas is only dominant up to $R\sim 0.6$).
 In contrast, the ICM of massive clusters is almost exclusively in the form of hot plasma and spatially extends up to $R\sim 1.2$. 
 One can thus interpret that behaviour by the fact that hot gas material inside massive clusters is mostly gravitationally heated whereas in galaxy groups the hot gas distribution might be governed by anisotropic accretion processes, and thus, it appears highly asymmetric. 
 In agreement with this interpretation, we found that considering all gas cells inside group and cluster halos, the anti-correlation between azimuthal symmetry of gas component and halo mass is removed. It means that only the hot plasma medium is strongly anisotropic inside galaxy groups.

Focusing on cluster peripheries from $1$ to $2$ $R_{200}$ in the bottom panel of Fig. \ref{fig:FIG_MASS}, we show that the anisotropic signatures of DM and WHIM are increasing with the cluster mass.
The small increase of DM azimuthal symmetric excess with halo mass, must be induced by the amount of filamentary structures which are expected to be more massive and more numerous around massive objects compared to low-mass ones \citep{Aragon2010,Codis2018,Sarron2019,Kraljic2020,Gouin2021}.
Regarding the azimuthal level of WHIM gas, there is a strong mass dependency: WHIM distribution around massive clusters is significantly more asymmetric than around low-mass groups.
In fact, we see in Fig. \ref{fig:FIG_V_R} that WHIM gas from $1$ to $2$ $R_{200}$ is rapidly infalling into massive clusters, whereas WHIM in galaxy group environments is slowly infalling and back-splashing.
In agreement with this dynamical picture, one can expect that WHIM gas is strongly asymmetric around massive objects because it is infalling along filamentary structures. This is consistent with \cite{Rost2021} who found that gas preferentially enters into massive clusters funneled by filaments.
In contrast, WHIM around groups is more isotropically distributed because it is rather accumulating (and ejected) at the group peripheries, and thus it must trace filaments more faintly.

To confirm if WHIM anisotropic distribution outside clusters is in general the result of filamentary pattern surrounding them, we show in Fig. \ref{fig:FIG_Connect} the azimuthal symmetric excess $\beta$ as a function of the connectivity of halos. As we can see, the number of cosmic filaments that are connected to clusters and the departure from spherical symmetry are significantly correlated for both the DM and the WHIM gas-phase.
In agreement with \cite{Galarraga2021} who found that WHIM gas-phase is strongly dominant in cosmic filaments, we conclude that the azimuthal distribution of WHIM tends to follow DM by tracing cosmic filamentary structures connected to clusters at their peripheries.

\begin{figure*}
    \centering
    \includegraphics[width=0.98\textwidth]{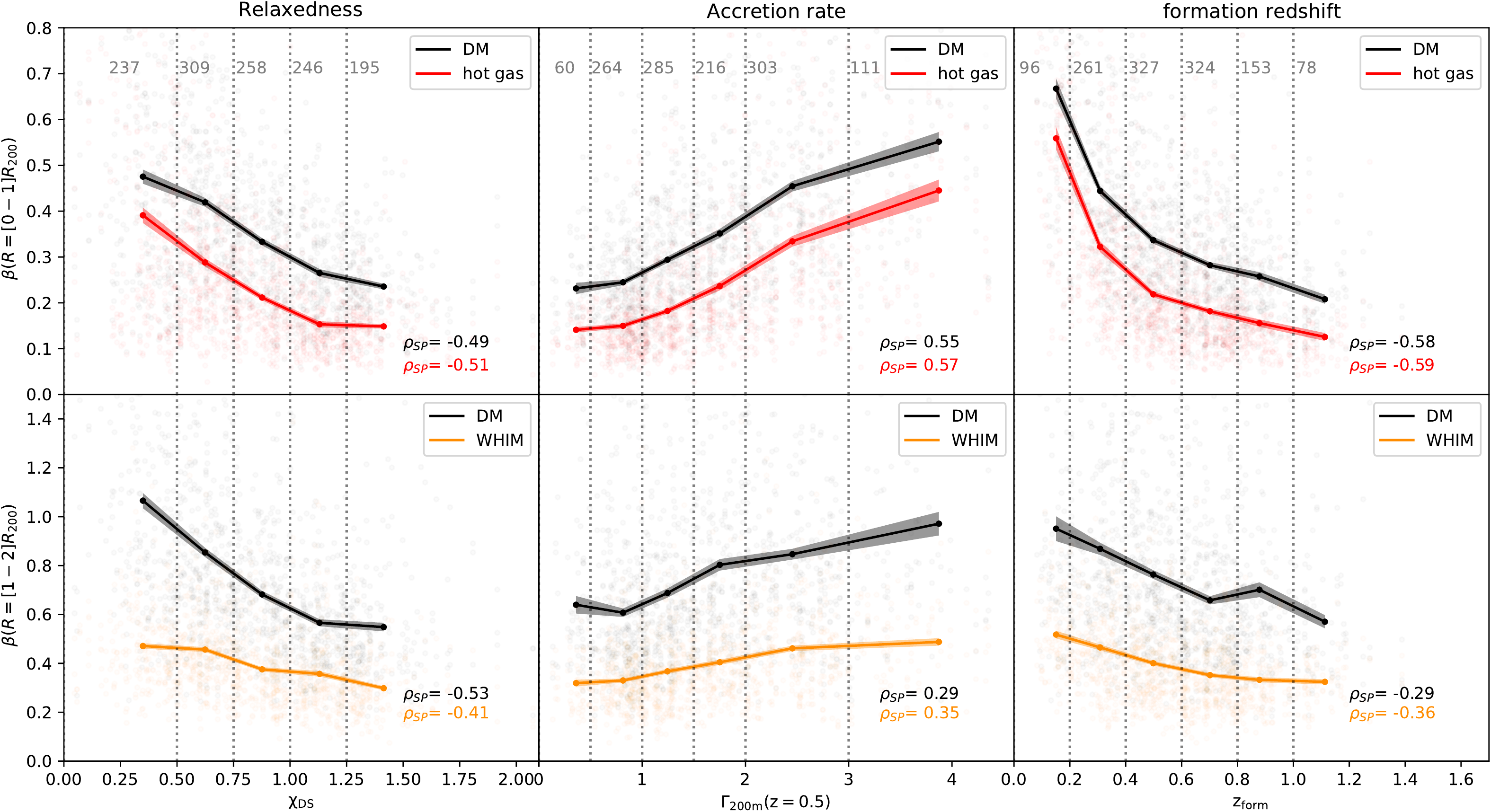}
    \caption{Distribution of the azimuthal symmetric excess $\beta$ (as defined in Eq. \ref{EQ:BETA_1to4}) computed inside clusters ($R<R_{200}$) in top panels, and at cluster peripheries in ($1<R [R_{200}]<2$) in bottom panels, as a function of different halo properties: level of relaxation on the left, mass accretion rate in the middle, and formation redshift in the right. The mean profiles of $\beta$ and their errors are shown in solid lines. The number of objects used to compute the average in each bin of x-axis (shown in gray dotted lines) is written on the top of the figures in gray}. 
    \label{fig:FIG_PHYSICAL}
\end{figure*}

\subsection{Accretion history dependency}

We have seen that the hot plasma traces well the halo properties of clusters, whereas the WHIM azimuthal distribution correlates with filamentary patterns connected to clusters. 
Given these findings, one can ask oneself whether the gas azimuthal distribution can also trace the overall cluster dynamics and its recent mass-assembly history, as well as DM. 
We refer here to Sect. \ref{SEC:CLUSTER}, for details on the definitions of the mass accretion rate $\Gamma$ (see Eq. \ref{eq:MAR}) and the formation redshift $z_{form}$ (see Eq. \ref{eq:zform}) which are two proxies of cluster mass-assembly history, and the relaxedness parameter $\chi_{DS}$ used to estimate cluster dynamical stat (see Eq. \ref{eq:DS}).

In Fig. \ref{fig:FIG_PHYSICAL}, we show the dependency of azimuthal symmetric excess $\beta$ with the different cluster properties: relaxedness (left panels), mass accretion rate (middle panels), and formation redshift (right panels). The azimuthal symmetric excess of hot gas and DM inside clusters ($R<R_{200}$) is presented in the top panels, whereas the bottom ones show the WHIM and DM anisotropic level at cluster peripheries ($1<R [R_{200}]<2$).

The departure from circular symmetry is fist analysed as a function of the relaxation level of clusters (left panels). We found that, the more circular the matter distribution ($\beta \longrightarrow 0$) the more dynamically relaxed the cluster halo  ($\chi_{DS}>1$).
Inside clusters, halo relaxedness and asymmetry are strongly correlated for both DM and hot gas, with $\rho_{SP}\sim 0.5$, showing that the azimuthal distribution of hot plasma is a powerful probe of cluster dynamics.
This result is in agreement with previous investigations on ICM asymmetry to probe cluster dynamical state \citep{Vazza2011,Eckert2012,Capalbo2021}.
Moreover, at cluster peripheries the azimuthal distribution of WHIM and DM also traces well the cluster relaxation level. This can be explained by the fact that the cluster dynamical state must be affected by the number of cosmic filaments connected to clusters, as shown in \cite{Gouin2021}.

Secondly, the influence of the cluster mass-assembly history on the azimuthal matter distribution is investigated by considering two variables, the mass accretion rate and the formation redshift of clusters in the middle and right panels, respectively.
Both cluster accretion history proxies are correlated with gas and DM azimuthal symmetry excess, in particular inside cluster halos, with $\rho_{SP}\sim 0.6$. 
The faster the cluster accretes material and the more recently formed, the more asymmetric its hot medium distribution.
This is coherent with the result of \cite{Chen2020} who show that the ellipticity of ICM  seems to be an imprint of the mass assembly history of clusters. 
Besides the ICM, we found that WHIM and DM inside the filamentary structure at the cluster periphery show a similar trend. 
Indeed, \cite{Gouin2021} have shown that the cosmic filaments connected to clusters affect cluster assembly history. 
Nevertheless, the correlation between the azimuthal symmetry and mass-assembly history decreases up to $\rho_{SP}\sim 0.3$ at cluster peripheries, suggesting that the mass distribution in the inner regions is more sensitive to the cluster past accretion history. 

\section{Discussion}

In this study, we have applied the azimuthal decomposition technique to the spatial distribution of simulated group and cluster gas phases defined using temperature-density selected gas cells projected in 2-D. 
We have shown that the 2-D azimuthal gas distribution is related to the 3-D cluster mass distribution from their shape to their connected filaments. Indeed, the correlation between azimuthal symmetric excess (in 2-D) and the connectivity of clusters shows that considering only main contribution of harmonic decomposition traces successfully the filamentary patterns \citep[see also][]{Valles2020}. However, we can expect that the 3-D spherical harmonics must be more correlated with 3-D ellipticity and 3-D connectivity than 2-D harmonics one, due to the small corrections provided by the full spherical harmonic decomposition.

The next step in the use of the multipolar analysis will be to apply 2-D multipole moment formalism on 2-D mock X-ray and SZ images to accurately evaluate the efficiency of such a technique on future data by taking into account observational effects such as finite angular resolution due to instrumental beam, noise, etc \citep[see e.g.][for mock X-ray images of clusters]{Green2019,DeLuca2020,Comparat2020O,Gianfagna2021}.  
This approach could be a powerful tool to highlight patterns in current and upcoming observations of the cosmic gas at the peripheries of galaxy clusters, such as those of X-COP \citep{XCOP}, Cluster Heritage \citep{Heritage}, and eROSITA \citep{eRosita,Comparat2022}. Indeed, eROSITA mission is expected to provide the necessary resolution to statistically probe connected gas filaments around clusters with the harmonic decomposition technique, as it has already allowed the discovery of long WHIM filament \citep{EROSITA_Reiprich2021,EROSITA_Angie2021,EROSITA_Bruggen2021,eRosita_Biffi2021}. 
Moreover, on much smaller scale than the galaxy clusters (at few $kpc$), the multipole decomposition might also be a powerful tool for other application such as quantifying the anisotropic distribution of circum galactic medium around black holes as predicted by \cite{Truong2021}, to constrain super massive black hole feedback from X-ray observations.

\section{Conclusions}

In this study, we have explored how gas and DM components are distributed in galaxy cluster environments from their cores to their connection to cosmic filaments. We focused on the matter distribution in $415$ galaxy clusters environments, defined as the regions extending up to $5 R_{200}$, extracted from the IllustrisTNG simulation.
Their gas phases, defined thanks to the temperature-density diagram, and their motions were first investigated as a function of the cluster-centric distance and the mass. 
This allowed us to identify the most relevant gas phases and radial apertures to further study the azimuthal distribution of the gas in clusters. By decomposing the matter distribution in harmonic space, we statistically explored the angular features, also called azimuthal symmetries, in gas and DM distributions around clusters.
Previous multipolar expansion of dark matter \citep{Gouin2017} and galaxy \citep{Gouin2020} distributions in cluster environments have revealed the usefulness of this approach to statistically probe the cosmic filamentary pattern outside clusters. 
In the present extension to the studies by \cite{Gouin2017} and \cite{Gouin2020}, we focused on gas distribution to explore how azimuthal symmetries of gas in different phases trace cluster physical and dynamical properties as well as the \textit{"reference"} dark matter distribution.
The radial and azimuthal gas distribution from cluster inner regions to their connection to filaments shows that:

\begin{itemize}

    \item[(1)] Galaxy cluster environments are key regions where the warm diffuse gas is infalling into clusters, and heated to become the hot plasma inside cluster cores \citep[as also discussed by][in the large cosmic web picture]{Martizzi2019}.
    In detail, the hot gas phase is dominant inside clusters, up to around $1 \, R_{200}$, and is well virialised with an equal balance between slow inflow and outflow motions. 
    In contrast, the gas outside clusters ($R>R_{200}$) is mainly in the form of WHIM and presents two distinct dynamical regimes: the accumulating gas at cluster peripheries from $\sim 1$ to $\sim 2 \ R_{200}$ with both slowly ejected and infalling gas motions, and the rapidly infalling gas for distance from the center $R>1.5 \ R_{200}$.
    These findings can be put in perspective of \cite{Rost2021}, who found that cosmic gas is fast infalling into clusters along filaments, and leaves the cluster outside filaments.
    
    \item[(2)] Groups and clusters show different radial transitions from hot to warm gas phases. 
     While galaxy clusters present an extended core of hot plasma up to $1.2\,R_{200}$, the galaxy groups contain different gas phases with a core of hot plasma (up to $0.6\,R_{200}$), a shell of warm dense gas (WCGM phase) and diffuse warm gas (WHIM phase) beyond $R>0.8\,R_{200}$. 
     Groups and clusters also show different WHIM motion at their outskirts. Whereas the WHIM gas-phase is mostly rapidly infalling outside clusters, it is rather accumulating and slowly infalling at the peripheries of groups.

    \item[(3)] The azimuthal symmetric features of gas and DM distributions are strongly correlated with distinct structural properties of the cluster halos.
    In particular, the dipolar symmetry reflects the halo center offset, the quadrupolar symmetry describes the halo ellipticity, and larger harmonic decomposition orders trace the amount of halo substructures. The azimuthal hot plasma distribution appears to follow well the DM one, tracing the structure of the cluster halo, even if the hot gas tends to be smoother and more circular than the DM distribution \citep[as expected from the ellipsoidal shape of DM and hot gas found in][]{Okabe2018}.  
    
    \item[(4)] Focusing on mass dependency, we found that the hot gas is more asymmetric inside galaxy groups than in clusters. We relate this to the fact that the hot gas does not represent the overall gas distribution inside groups, but is rather concentrated in group cores and must be subject to distortion by anisotropic accretion processes. In contrast, the ICM of massive clusters is almost only composed of gravitationally heated gas inside $R_{200}$ (at 93\%), and thus, it can explain its more isotropic shape. 
    
    \item[(5)] At cluster peripheries, the WHIM and DM azimuthal symmetries increase with cluster mass, in agreement with the expected increase of the cosmic filamentary signature with mass in harmonic space \citep{Gouin2020}. Notice that WHIM at peripheries of massive clusters better traces cosmic filament patterns than in the low-mass clusters and groups. This can be explained according to our dynamical analysis of WHIM gas, and to results from \cite{Rost2021}, by the fact that WHIM around groups is both slowly infalling in filaments and outflowing from groups out of filaments. 
    
    \item[(6)] At cluster peripheries, the asymmetric signatures of WHIM and DM distributions increase significantly with the number of connected filaments, showing that the matter azimuthal symmetric excess in cluster infalling regions (from $1$ to $2\,R_{200}$) is an imprint of cluster filamentary patterns. Therefore, the WHIM gas phase, as it follows the DM distribution, appears to trace well connections to the cosmic filaments. We also confirm that gas in filaments outside clusters is mostly in the form of WHIM phase, in agreement with \cite{Galarraga2021}.
    
    \item[(7)] We found that the gas azimuthal distribution is affected by the past assembly history of clusters and that it is a good tracer of its current dynamical state, as good as the \textit{"reference"} DM distribution. 
    In agreement with previous ICM investigations in simulations \citep[see e.g.][]{Vazza2011}, we confirm that, by using this azimuthal decomposition technique, departures from circular symmetry in hot gas distribution are stronger for more dynamically unrelaxed, faster-accreting and later formed clusters. Beyond these findings, we show that these relations between gas distribution and cluster properties can be extended to the warm diffuse gas phase located in cosmic filaments at cluster peripheries. 
    
\end{itemize}

By probing both the radial and azimuthal gas distribution in galaxy cluster environments, we describe the radial transition from hot gas, dominant in the inner regions, to WHIM gas, dominant at the peripheries, during their infall along cosmic filaments. We found that gas properties and it distribution are strongly affected by cluster environments up to about $2 \, R_{200}$. Beyond radial distance of $4 \, R_{200}$, the cluster environments no more impact gas azimuthal distribution. 

Moreover, by using this novel reliable technique based on multipolar decomposition, we statistically probed azimuthal symmetric features in the gas distribution up to the connection with filaments at cluster peripheries. Our study constitutes a first attempt to statistically explore azimuthal gas distribution in 2-D projected gas maps in its main phases, comparable to gas observations in galaxy cluster environments \citep{Eckert2015}. 
In real data, this approach might require high angular resolution \citep[similarly to][]{Valles2021,Capalbo2021}, and could be a powerful tool to highlight patterns in current and future observations of the cosmic gas at the peripheries of galaxy clusters  \citep{XCOP,Athena,XRISM,Heritage,Simionescu2021}.

\begin{acknowledgements}
We thank the anonymous referee for his/her report and helpful comments and suggestions. This research was supported by the funding for the ByoPiC project from the European Research Council (ERC) under the European Union’s Horizon 2020 research and innovation program grant agreement ERC-2015-AdG 695561 (ByoPiC, https://byopic.eu). The authors thank the very useful comments and discussions with all the members of the ByoPiC team. We thank the IllustrisTNG collaboration for providing free access to the data used in this work.
CG is supported by a KIAS Individual Grant (PG085001) at Korea Institute for Advanced Study (KIAS). She is grateful to Professor Changbom Park for stimulating discussions.
\end{acknowledgements}

\bibliographystyle{aa}

\begin{thebibliography}{127}
\expandafter\ifx\csname natexlab\endcsname\relax\def\natexlab#1{#1}\fi

\bibitem[{{Akamatsu} {et~al.}(2017){Akamatsu}, {Fujita}, {Akahori}, {Ishisaki},
  {Hayashida}, {Hoshino}, {Mernier}, {Yoshikawa}, {Sato}, \&
  {Kaastra}}]{Akamatsu2017}
{Akamatsu}, H., {Fujita}, Y., {Akahori}, T., {et~al.} 2017, \aap, 606, A1

\bibitem[{{Anand} {et~al.}(2022){Anand}, {Kauffmann}, \& {Nelson}}]{Anand2022}
{Anand}, A., {Kauffmann}, G., \& {Nelson}, D. 2022, arXiv e-prints,
  arXiv:2201.07811

\bibitem[{{Angelinelli} {et~al.}(2021){Angelinelli}, {Ettori}, {Vazza}, \&
  {Jones}}]{Angelinelli2021}
{Angelinelli}, M., {Ettori}, S., {Vazza}, F., \& {Jones}, T.~W. 2021, arXiv
  e-prints, arXiv:2102.01096

\bibitem[{{Angelinelli} {et~al.}(2020){Angelinelli}, {Vazza}, {Giocoli},
  {Ettori}, {Jones}, {Brunetti}, {Br{\"u}ggen}, \& {Eckert}}]{Angelinelli2020}
{Angelinelli}, M., {Vazza}, F., {Giocoli}, C., {et~al.} 2020, \mnras, 495, 864

\bibitem[{{Ansarifard} {et~al.}(2020){Ansarifard}, {Rasia}, {Biffi}, {Borgani},
  {Cui}, {De Petris}, {Dolag}, {Ettori}, {Movahed}, {Murante}, \&
  {Yepes}}]{Ansarifard2020}
{Ansarifard}, S., {Rasia}, E., {Biffi}, V., {et~al.} 2020, \aap, 634, A113

\bibitem[{{Arag{\'o}n-Calvo} {et~al.}(2010{\natexlab{a}}){Arag{\'o}n-Calvo},
  {Platen}, {van de Weygaert}, \& {Szalay}}]{Aragon2010_spineweb}
{Arag{\'o}n-Calvo}, M.~A., {Platen}, E., {van de Weygaert}, R., \& {Szalay},
  A.~S. 2010{\natexlab{a}}, \apj, 723, 364

\bibitem[{{Arag{\'o}n-Calvo} {et~al.}(2010{\natexlab{b}}){Arag{\'o}n-Calvo},
  {van de Weygaert}, \& {Jones}}]{Aragon2010}
{Arag{\'o}n-Calvo}, M.~A., {van de Weygaert}, R., \& {Jones}, B. J.~T.
  2010{\natexlab{b}}, \mnras, 408, 2163

\bibitem[{{Artale} {et~al.}(2022){Artale}, {Haider}, {Montero-Dorta},
  {Vogelsberger}, {Martizzi}, {Torrey}, {Bird}, {Hernquist}, \&
  {Marinacci}}]{Artale2022}
{Artale}, M.~C., {Haider}, M., {Montero-Dorta}, A.~D., {et~al.} 2022, \mnras,
  510, 399

\bibitem[{{Arthur} {et~al.}(2019){Arthur}, {Pearce}, {Gray}, {Knebe}, {Cui},
  {Elahi}, {Power}, {Yepes}, {Arth}, {De Petris}, {Dolag}, {Garratt-Smithson},
  {Old}, {Rasia}, \& {Stevens}}]{Arthur2019}
{Arthur}, J., {Pearce}, F.~R., {Gray}, M.~E., {et~al.} 2019, \mnras, 484, 3968

\bibitem[{{Barcons} {et~al.}(2017){Barcons}, {Barret}, {Decourchelle}, {den
  Herder}, {Fabian}, {Matsumoto}, {Lumb}, {Nandra}, {Piro}, {Smith}, \&
  {Willingale}}]{Athena}
{Barcons}, X., {Barret}, D., {Decourchelle}, A., {et~al.} 2017, Astronomische
  Nachrichten, 338, 153

\bibitem[{{Biffi} {et~al.}(2021){Biffi}, {Dolag}, {Reiprich}, {Veronica},
  {Ramos-Ceja}, {Bulbul}, {Ota}, \& {Ghirardini}}]{eRosita_Biffi2021}
{Biffi}, V., {Dolag}, K., {Reiprich}, T.~H., {et~al.} 2021, arXiv e-prints,
  arXiv:2106.14542

\bibitem[{{Bond} \& {Myers}(1996)}]{Bond1996}
{Bond}, J.~R. \& {Myers}, S.~T. 1996, \apjs, 103, 63

\bibitem[{{Bonjean} {et~al.}(2018){Bonjean}, {Aghanim}, {Salom{\'e}},
  {Douspis}, \& {Beelen}}]{Bonjean2018}
{Bonjean}, V., {Aghanim}, N., {Salom{\'e}}, P., {Douspis}, M., \& {Beelen}, A.
  2018, \aap, 609, A49

\bibitem[{{Bonnaire} {et~al.}(2020){Bonnaire}, {Aghanim}, {Decelle}, \&
  {Douspis}}]{Bonnaire2020}
{Bonnaire}, T., {Aghanim}, N., {Decelle}, A., \& {Douspis}, M. 2020, \aap, 637,
  A18

\bibitem[{{Bonnaire} {et~al.}(2021){Bonnaire}, {Decelle}, \&
  {Aghanim}}]{Bonnaire2021}
{Bonnaire}, T., {Decelle}, A., \& {Aghanim}, N. 2021, arXiv e-prints,
  arXiv:2106.09035

\bibitem[{{Br{\"u}ggen} {et~al.}(2021){Br{\"u}ggen}, {Reiprich}, {Bulbul},
  {Koribalski}, {Andernach}, {Rudnick}, {Hoang}, {Wilber}, {Duchesne},
  {Veronica}, {Pacaud}, {Hopkins}, {Norris}, {Johnston-Hollitt}, {Brown},
  {Bonafede}, {Brunetti}, {Collier}, {Sanders}, {Vardoulaki}, {Venturi},
  {Kapinska}, \& {Marvil}}]{EROSITA_Bruggen2021}
{Br{\"u}ggen}, M., {Reiprich}, T.~H., {Bulbul}, E., {et~al.} 2021, \aap, 647,
  A3

\bibitem[{{Capalbo} {et~al.}(2021){Capalbo}, {De Petris}, {De Luca}, {Cui},
  {Yepes}, {Knebe}, \& {Rasia}}]{Capalbo2021}
{Capalbo}, V., {De Petris}, M., {De Luca}, F., {et~al.} 2021, \mnras, 503, 6155

\bibitem[{{Cautun} {et~al.}(2013){Cautun}, {van de Weygaert}, \&
  {Jones}}]{Cautun2013}
{Cautun}, M., {van de Weygaert}, R., \& {Jones}, B. J.~T. 2013, \mnras, 429,
  1286

\bibitem[{{Cautun} {et~al.}(2014){Cautun}, {van de Weygaert}, {Jones}, \&
  {Frenk}}]{Cautun2014}
{Cautun}, M., {van de Weygaert}, R., {Jones}, B. J.~T., \& {Frenk}, C.~S. 2014,
  \mnras, 441, 2923

\bibitem[{{Cen} \& {Ostriker}(1999)}]{Cen1999}
{Cen}, R. \& {Ostriker}, J.~P. 1999, \apj, 514, 1

\bibitem[{{Cen} \& {Ostriker}(2006)}]{Cen2006}
{Cen}, R. \& {Ostriker}, J.~P. 2006, \apj, 650, 560

\bibitem[{{Chen} {et~al.}(2019){Chen}, {Avestruz}, {Kravtsov}, {Lau}, \&
  {Nagai}}]{Chen2019}
{Chen}, H., {Avestruz}, C., {Kravtsov}, A.~V., {Lau}, E.~T., \& {Nagai}, D.
  2019, arXiv e-prints, arXiv:1903.08662

\bibitem[{{Chen} {et~al.}(2020){Chen}, {Mo}, {Li}, {Wang}, {Yang}, {Zhang}, \&
  {Wang}}]{Chen2020}
{Chen}, Y., {Mo}, H.~J., {Li}, C., {et~al.} 2020, \apj, 899, 81

\bibitem[{{CHEX-MATE Collaboration} {et~al.}(2021){CHEX-MATE Collaboration},
  {Arnaud}, {Ettori}, {Pratt}, {Rossetti}, {Eckert}, {Gastaldello}, {Gavazzi},
  {Kay}, {Lovisari}, {Maughan}, {Pointecouteau}, {Sereno}, {Bartalucci},
  {Bonafede}, {Bourdin}, {Cassano}, {Duffy}, {Iqbal}, {Maurogordato}, {Rasia},
  {Sayers}, {Andrade-Santos}, {Aussel}, {Barnes}, {Barrena}, {Borgani},
  {Burkutean}, {Clerc}, {Corasaniti}, {Cuillandre}, {De Grandi}, {De Petris},
  {Dolag}, {Donahue}, {Ferragamo}, {Gaspari}, {Ghizzardi}, {Gitti}, {Haines},
  {Jauzac}, {Johnston-Hollitt}, {Jones}, {K{\'e}ruzor{\'e}}, {LeBrun}, {Mayet},
  {Mazzotta}, {Melin}, {Molendi}, {Nonino}, {Okabe}, {Paltani}, {Perotto},
  {Pires}, {Radovich}, {Rubino-Martin}, {Salvati}, {Saro}, {Sartoris},
  {Schellenberger}, {Streblyanska}, {Tarr{\'\i}o}, {Tozzi}, {Umetsu}, {van der
  Burg}, {Vazza}, {Venturi}, {Yepes}, \& {Zarattini}}]{Heritage}
{CHEX-MATE Collaboration}, {Arnaud}, M., {Ettori}, S., {et~al.} 2021, \aap,
  650, A104

\bibitem[{{Christiansen} {et~al.}(2020){Christiansen}, {Dav{\'e}}, {Sorini}, \&
  {Angl{\'e}s-Alc{\'a}zar}}]{Christiansen2020}
{Christiansen}, J.~F., {Dav{\'e}}, R., {Sorini}, D., \&
  {Angl{\'e}s-Alc{\'a}zar}, D. 2020, \mnras, 499, 2617

\bibitem[{{Cialone} {et~al.}(2018){Cialone}, {De Petris}, {Sembolini}, {Yepes},
  {Baldi}, \& {Rasia}}]{Cialone2018}
{Cialone}, G., {De Petris}, M., {Sembolini}, F., {et~al.} 2018, \mnras, 477,
  139

\bibitem[{{Clampitt} \& {Jain}(2016)}]{Clampitt2016B}
{Clampitt}, J. \& {Jain}, B. 2016, \mnras, 457, 4135

\bibitem[{{Codis} {et~al.}(2017){Codis}, {Gavazzi}, {Pichon}, \&
  {Gouin}}]{Codis2017}
{Codis}, S., {Gavazzi}, R., {Pichon}, C., \& {Gouin}, C. 2017, \aap, 605, A80

\bibitem[{{Codis} {et~al.}(2015){Codis}, {Pichon}, \& {Pogosyan}}]{Codis2015}
{Codis}, S., {Pichon}, C., \& {Pogosyan}, D. 2015, \mnras, 452, 3369

\bibitem[{{Codis} {et~al.}(2018){Codis}, {Pogosyan}, \& {Pichon}}]{Codis2018}
{Codis}, S., {Pogosyan}, D., \& {Pichon}, C. 2018, \mnras, 479, 973

\bibitem[{{Cole} \& {Lacey}(1996)}]{Cole1996}
{Cole}, S. \& {Lacey}, C. 1996, \mnras, 281, 716

\bibitem[{{Comparat} {et~al.}(2020){Comparat}, {Eckert}, {Finoguenov},
  {Schmidt}, {Sanders}, {Nagai}, {Lau}, {K{\"a}}, {fer}, {Pacaud}, {Clerc},
  {Reiprich}, {Bulbul}, {Chitham}, {Chiang}, {Ghirardini}, {Gonzalez-Perez},
  {Gozaliasl}, {Fitzpatrick}, {Klypin}, {Merloni}, {Nandra}, {Liu}, {Prada},
  {Ramos-Ceja}, {Salvato}, {Seppi}, {Tempel}, \& {Yepes}}]{Comparat2020O}
{Comparat}, J., {Eckert}, D., {Finoguenov}, A., {et~al.} 2020, The Open Journal
  of Astrophysics, 3, 13

\bibitem[{{Comparat} {et~al.}(2022){Comparat}, {Truong}, {Merloni},
  {Pillepich}, {Ponti}, {Driver}, {Bellstedt}, {Liske}, {Aird}, {Br{\"u}ggen},
  {Bulbul}, {Davies}, {Gonz{\'a}lez Villalba}, {Georgakakis}, {Haberl}, {Liu},
  {Maitra}, {Nandra}, {Popesso}, {Predehl}, {Robotham}, {Salvato}, {Thorne}, \&
  {Zhang}}]{Comparat2022}
{Comparat}, J., {Truong}, N., {Merloni}, A., {et~al.} 2022, arXiv e-prints,
  arXiv:2201.05169

\bibitem[{{Cui} {et~al.}(2017){Cui}, {Power}, {Borgani}, {Knebe}, {Lewis},
  {Murante}, \& {Poole}}]{Cui2017}
{Cui}, W., {Power}, C., {Borgani}, S., {et~al.} 2017, \mnras, 464, 2502

\bibitem[{{Dacunha} {et~al.}(2021){Dacunha}, {Belyakov}, {Adhikari}, {Shin},
  {Goldstein}, \& {Jain}}]{Dacunha2021}
{Dacunha}, T., {Belyakov}, M., {Adhikari}, S., {et~al.} 2021, arXiv e-prints,
  arXiv:2111.06499

\bibitem[{{Darragh-Ford} {et~al.}(2019){Darragh-Ford}, {Laigle}, {Gozaliasl},
  {Pichon}, {Devriendt}, {Slyz}, {Arnouts}, {Dubois}, {Finoguenov},
  {Griffiths}, {Kraljic}, {Pan}, {Peirani}, \& {Sarron}}]{Darragh2019}
{Darragh-Ford}, E., {Laigle}, C., {Gozaliasl}, G., {et~al.} 2019, arXiv
  e-prints, arXiv:1904.11859

\bibitem[{{Dav{\'e}} {et~al.}(2001){Dav{\'e}}, {Cen}, {Ostriker}, {Bryan},
  {Hernquist}, {Katz}, {Weinberg}, {Norman}, \& {O'Shea}}]{Dave2001_WHIM}
{Dav{\'e}}, R., {Cen}, R., {Ostriker}, J.~P., {et~al.} 2001, \apj, 552, 473

\bibitem[{{Davis} {et~al.}(1985){Davis}, {Efstathiou}, {Frenk}, \&
  {White}}]{Davis1985}
{Davis}, M., {Efstathiou}, G., {Frenk}, C.~S., \& {White}, S.~D.~M. 1985, \apj,
  292, 371

\bibitem[{{de Graaff} {et~al.}(2019){de Graaff}, {Cai}, {Heymans}, \&
  {Peacock}}]{deGraaff2019}
{de Graaff}, A., {Cai}, Y.-C., {Heymans}, C., \& {Peacock}, J.~A. 2019, \aap,
  624, A48

\bibitem[{{de Lapparent} {et~al.}(1986){de Lapparent}, {Geller}, \&
  {Huchra}}]{deLapparent1986}
{de Lapparent}, V., {Geller}, M.~J., \& {Huchra}, J.~P. 1986, \apjl, 302, L1

\bibitem[{{De Luca} {et~al.}(2020){De Luca}, {De Petris}, {Yepes}, {Cui},
  {Knebe}, \& {Rasia}}]{DeLuca2020}
{De Luca}, F., {De Petris}, M., {Yepes}, G., {et~al.} 2020, arXiv e-prints,
  arXiv:2011.09002

\bibitem[{{Despali} {et~al.}(2017){Despali}, {Giocoli}, {Bonamigo}, {Limousin},
  \& {Tormen}}]{Despali2017}
{Despali}, G., {Giocoli}, C., {Bonamigo}, M., {Limousin}, M., \& {Tormen}, G.
  2017, \mnras, 466, 181

\bibitem[{{Despali} {et~al.}(2014){Despali}, {Giocoli}, \&
  {Tormen}}]{Despali2014}
{Despali}, G., {Giocoli}, C., \& {Tormen}, G. 2014, \mnras, 443, 3208

\bibitem[{{Diemer} \& {Kravtsov}(2014)}]{Diemer2014}
{Diemer}, B. \& {Kravtsov}, A.~V. 2014, \apj, 789, 1

\bibitem[{{Diemer} {et~al.}(2013){Diemer}, {More}, \& {Kravtsov}}]{Diemer2013b}
{Diemer}, B., {More}, S., \& {Kravtsov}, A.~V. 2013, \apj, 766, 25

\bibitem[{{Dietrich} {et~al.}(2005){Dietrich}, {Schneider}, {Clowe},
  {Romano-D{\'\i}az}, \& {Kerp}}]{Dietrich2005}
{Dietrich}, J.~P., {Schneider}, P., {Clowe}, D., {Romano-D{\'\i}az}, E., \&
  {Kerp}, J. 2005, \aap, 440, 453

\bibitem[{{Donahue} {et~al.}(2016){Donahue}, {Ettori}, {Rasia}, {Sayers},
  {Zitrin}, {Meneghetti}, {Voit}, {Golwala}, {Czakon}, {Yepes}, {Baldi},
  {Koekemoer}, \& {Postman}}]{Donahue2016}
{Donahue}, M., {Ettori}, S., {Rasia}, E., {et~al.} 2016, \apj, 819, 36

\bibitem[{{Eckert} {et~al.}(2015){Eckert}, {Jauzac}, {Shan}, {Kneib}, {Erben},
  {Israel}, {Jullo}, {Klein}, {Massey}, {Richard}, \& {Tchernin}}]{Eckert2015}
{Eckert}, D., {Jauzac}, M., {Shan}, H., {et~al.} 2015, \nat, 528, 105

\bibitem[{{Eckert} {et~al.}(2012){Eckert}, {Vazza}, {Ettori}, {Molendi},
  {Nagai}, {Lau}, {Roncarelli}, {Rossetti}, {Snowden}, \&
  {Gastaldello}}]{Eckert2012}
{Eckert}, D., {Vazza}, F., {Ettori}, S., {et~al.} 2012, \aap, 541, A57

\bibitem[{{Einasto} {et~al.}(2020){Einasto}, {Deshev}, {Tenjes},
  {Hein{\"a}m{\"a}ki}, {Tempel}, {Juhan Liivam{\"a}gi}, {Einasto}, {Lietzen},
  {Tuvikene}, \& {Chon}}]{Einasto2020}
{Einasto}, M., {Deshev}, B., {Tenjes}, P., {et~al.} 2020, \aap, 641, A172

\bibitem[{{Gal{\'a}rraga-Espinosa} {et~al.}(2020){Gal{\'a}rraga-Espinosa},
  {Aghanim}, {Langer}, {Gouin}, \& {Malavasi}}]{Galarraga2020}
{Gal{\'a}rraga-Espinosa}, D., {Aghanim}, N., {Langer}, M., {Gouin}, C., \&
  {Malavasi}, N. 2020, \aap, 641, A173

\bibitem[{{Gal{\'a}rraga-Espinosa}
  {et~al.}(2021{\natexlab{a}}){Gal{\'a}rraga-Espinosa}, {Aghanim}, {Langer}, \&
  {Tanimura}}]{Galarraga2021}
{Gal{\'a}rraga-Espinosa}, D., {Aghanim}, N., {Langer}, M., \& {Tanimura}, H.
  2021{\natexlab{a}}, \aap, 649, A117

\bibitem[{{Gal{\'a}rraga-Espinosa}
  {et~al.}(2021{\natexlab{b}}){Gal{\'a}rraga-Espinosa}, {Langer}, \&
  {Aghanim}}]{Galarraga2021b}
{Gal{\'a}rraga-Espinosa}, D., {Langer}, M., \& {Aghanim}, N.
  2021{\natexlab{b}}, arXiv e-prints, arXiv:2109.06198

\bibitem[{{Gheller} \& {Vazza}(2019)}]{Gheller2019}
{Gheller}, C. \& {Vazza}, F. 2019, \mnras, 486, 981

\bibitem[{{Gianfagna} {et~al.}(2021){Gianfagna}, {De Petris}, {Yepes}, {De
  Luca}, {Sembolini}, {Cui}, {Biffi}, {K{\'e}ruzor{\'e}},
  {Mac{\'\i}as-P{\'e}rez}, {Mayet}, {Perotto}, {Rasia}, \&
  {Ruppin}}]{Gianfagna2021}
{Gianfagna}, G., {De Petris}, M., {Yepes}, G., {et~al.} 2021, \mnras, 502, 5115

\bibitem[{{Gouin} {et~al.}(2020){Gouin}, {Aghanim}, {Bonjean}, \&
  {Douspis}}]{Gouin2020}
{Gouin}, C., {Aghanim}, N., {Bonjean}, V., \& {Douspis}, M. 2020, \aap, 635,
  A195

\bibitem[{{Gouin} {et~al.}(2021){Gouin}, {Bonnaire}, \& {Aghanim}}]{Gouin2021}
{Gouin}, C., {Bonnaire}, T., \& {Aghanim}, N. 2021, \aap, 651, A56

\bibitem[{{Gouin} {et~al.}(2017){Gouin}, {Gavazzi}, {Codis}, {Pichon},
  {Peirani}, \& {Dubois}}]{Gouin2017}
{Gouin}, C., {Gavazzi}, R., {Codis}, S., {et~al.} 2017, \aap, 605, A27

\bibitem[{{Green} {et~al.}(2019){Green}, {Ntampaka}, {Nagai}, {Lovisari},
  {Dolag}, {Eckert}, \& {ZuHone}}]{Green2019}
{Green}, S.~B., {Ntampaka}, M., {Nagai}, D., {et~al.} 2019, \apj, 884, 33

\bibitem[{{Haggar} {et~al.}(2020){Haggar}, {Gray}, {Pearce}, {Knebe}, {Cui},
  {Mostoghiu}, \& {Yepes}}]{Haggar2020}
{Haggar}, R., {Gray}, M.~E., {Pearce}, F.~R., {et~al.} 2020, \mnras, 492, 6074

\bibitem[{{Hahn}(2016)}]{Hahn2016_review}
{Hahn}, O. 2016, in The Zeldovich Universe: Genesis and Growth of the Cosmic
  Web, ed. R.~{van de Weygaert}, S.~{Shandarin}, E.~{Saar}, \& J.~{Einasto},
  Vol. 308, 87--96

\bibitem[{{Hahn} {et~al.}(2007){Hahn}, {Carollo}, {Porciani}, \&
  {Dekel}}]{Hahn200_TWEB}
{Hahn}, O., {Carollo}, C.~M., {Porciani}, C., \& {Dekel}, A. 2007, \mnras, 381,
  41

\bibitem[{{Jing} \& {Suto}(2002)}]{Jing2002}
{Jing}, Y.~P. \& {Suto}, Y. 2002, \apj, 574, 538

\bibitem[{{Kraljic} {et~al.}(2020){Kraljic}, {Pichon}, {Codis}, {Laigle},
  {Dav{\'e}}, {Dubois}, {Hwang}, {Pogosyan}, {Arnouts}, {Devriendt}, {Musso},
  {Peirani}, {Slyz}, \& {Treyer}}]{Kraljic2020}
{Kraljic}, K., {Pichon}, C., {Codis}, S., {et~al.} 2020, \mnras, 491, 4294

\bibitem[{{Kuchner} {et~al.}(2020){Kuchner}, {Arag{\'o}n-Salamanca}, {Pearce},
  {Gray}, {Rost}, {Mu}, {Welker}, {Cui}, {Haggar}, {Laigle}, {Knebe},
  {Kraljic}, {Sarron}, \& {Yepes}}]{Kuchner2020}
{Kuchner}, U., {Arag{\'o}n-Salamanca}, A., {Pearce}, F.~R., {et~al.} 2020,
  \mnras, 494, 5473

\bibitem[{{Lee} {et~al.}(2021){Lee}, {Hwang}, \& {Song}}]{Lee2021_MGII}
{Lee}, J.~C., {Hwang}, H.~S., \& {Song}, H. 2021, \mnras, 503, 4309

\bibitem[{{Mahajan} {et~al.}(2018){Mahajan}, {Singh}, \&
  {Shobhana}}]{Mahajan2018}
{Mahajan}, S., {Singh}, A., \& {Shobhana}, D. 2018, \mnras, 478, 4336

\bibitem[{{Malavasi} {et~al.}(2020){Malavasi}, {Aghanim}, {Tanimura},
  {Bonjean}, \& {Douspis}}]{Malavasi2020}
{Malavasi}, N., {Aghanim}, N., {Tanimura}, H., {Bonjean}, V., \& {Douspis}, M.
  2020, \aap, 634, A30

\bibitem[{{Martizzi} {et~al.}(2019){Martizzi}, {Vogelsberger}, {Artale},
  {Haider}, {Torrey}, {Marinacci}, {Nelson}, {Pillepich}, {Weinberger},
  {Hernquist}, {Naiman}, \& {Springel}}]{Martizzi2019}
{Martizzi}, D., {Vogelsberger}, M., {Artale}, M.~C., {et~al.} 2019, \mnras,
  486, 3766

\bibitem[{{Mead} {et~al.}(2010){Mead}, {King}, \& {McCarthy}}]{Mead2010}
{Mead}, J. M.~G., {King}, L.~J., \& {McCarthy}, I.~G. 2010, \mnras, 401, 2257

\bibitem[{{Mishra} \& {Muzahid}(2022)}]{Mishra2022}
{Mishra}, S. \& {Muzahid}, S. 2022, arXiv e-prints, arXiv:2201.08545

\bibitem[{{Mohr} {et~al.}(1993){Mohr}, {Fabricant}, \& {Geller}}]{Mohr1993}
{Mohr}, J.~J., {Fabricant}, D.~G., \& {Geller}, M.~J. 1993, \apj, 413, 492

\bibitem[{{More} {et~al.}(2015){More}, {Diemer}, \& {Kravtsov}}]{More2015}
{More}, S., {Diemer}, B., \& {Kravtsov}, A.~V. 2015, \apj, 810, 36

\bibitem[{{Mostoghiu} {et~al.}(2021){Mostoghiu}, {Arthur}, {Pearce}, {Gray},
  {Knebe}, {Cui}, {Welker}, {Cora}, {Murante}, {Dolag}, \&
  {Yepes}}]{Mostoghiu2021}
{Mostoghiu}, R., {Arthur}, J., {Pearce}, F.~R., {et~al.} 2021, \mnras, 501,
  5029

\bibitem[{{Mostoghiu} {et~al.}(2019){Mostoghiu}, {Knebe}, {Cui}, {Pearce},
  {Yepes}, {Power}, {Dave}, \& {Arth}}]{Mostoghiu2019}
{Mostoghiu}, R., {Knebe}, A., {Cui}, W., {et~al.} 2019, \mnras, 483, 3390

\bibitem[{{Nelson} {et~al.}(2019){Nelson}, {Springel}, {Pillepich},
  {Rodriguez-Gomez}, {Torrey}, {Genel}, {Vogelsberger}, {Pakmor}, {Marinacci},
  {Weinberger}, {Kelley}, {Lovell}, {Diemer}, \& {Hernquist}}]{ILLUSTRIS_TNG}
{Nelson}, D., {Springel}, V., {Pillepich}, A., {et~al.} 2019, Computational
  Astrophysics and Cosmology, 6, 2

\bibitem[{{Nicastro} {et~al.}(2018){Nicastro}, {Kaastra}, {Krongold},
  {Borgani}, {Branchini}, {Cen}, {Dadina}, {Danforth}, {Elvis}, {Fiore},
  {Gupta}, {Mathur}, {Mayya}, {Paerels}, {Piro}, {Rosa-Gonzalez}, {Schaye},
  {Shull}, {Torres-Zafra}, {Wijers}, \& {Zappacosta}}]{Nicastro2018}
{Nicastro}, F., {Kaastra}, J., {Krongold}, Y., {et~al.} 2018, \nat, 558, 406

\bibitem[{{Okabe} {et~al.}(2018){Okabe}, {Nishimichi}, {Oguri}, {Peirani},
  {Kitayama}, {Sasaki}, \& {Suto}}]{Okabe2018}
{Okabe}, T., {Nishimichi}, T., {Oguri}, M., {et~al.} 2018, \mnras, 478, 1141

\bibitem[{{Oman} {et~al.}(2013){Oman}, {Hudson}, \& {Behroozi}}]{Oman2013}
{Oman}, K.~A., {Hudson}, M.~J., \& {Behroozi}, P.~S. 2013, \mnras, 431, 2307

\bibitem[{{Pereyra} {et~al.}(2020){Pereyra}, {Sgr{\'o}}, {Merch{\'a}n},
  {Stasyszyn}, \& {Paz}}]{Pereyra2020}
{Pereyra}, L.~A., {Sgr{\'o}}, M.~A., {Merch{\'a}n}, M.~E., {Stasyszyn}, F.~A.,
  \& {Paz}, D.~J. 2020, \mnras, 499, 4876

\bibitem[{{Pichon} {et~al.}(2010){Pichon}, {Gay}, {Pogosyan}, {Prunet},
  {Sousbie}, {Colombi}, {Slyz}, \& {Devriendt}}]{Pichon2010}
{Pichon}, C., {Gay}, C., {Pogosyan}, D., {et~al.} 2010, in American Institute
  of Physics Conference Series, Vol. 1241, American Institute of Physics
  Conference Series, ed. J.-M. {Alimi} \& A.~{Fu{\"o}zfa}, 1108--1117

\bibitem[{{Planck Collaboration} {et~al.}(2013){Planck Collaboration}, {Ade},
  {Aghanim}, {Arnaud}, {Ashdown}, {Atrio-Barandela}, {Aumont}, {Baccigalupi},
  {Balbi}, {Banday}, {Barreiro}, {Battaner}, {Benabed}, {Beno{\^\i}t},
  {Bernard}, {Bersanelli}, {Bhatia}, {Bikmaev}, {B{\"o}hringer}, {Bonaldi},
  {Bond}, {Borrill}, {Bouchet}, {Bourdin}, {Burenin}, {Burigana}, {Cabella},
  {Cardoso}, {Castex}, {Catalano}, {Cay{\'o}n}, {Chamballu}, {Chary}, {Chiang},
  {Chon}, {Christensen}, {Clements}, {Colafrancesco}, {Colombo}, {Comis},
  {Coulais}, {Crill}, {Cuttaia}, {Danese}, {Davis}, {de Bernardis}, {de
  Gasperis}, {de Zotti}, {Delabrouille}, {D{\'e}mocl{\`e}s}, {D{\'e}sert},
  {Diego}, {Dolag}, {Dole}, {Donzelli}, {Dor{\'e}}, {D{\"o}rl}, {Douspis},
  {Dupac}, {Efstathiou}, {En{\ss}lin}, {Eriksen}, {Finelli}, {Flores-Cacho},
  {Forni}, {Frailis}, {Franceschi}, {Frommert}, {Ganga}, {G{\'e}nova-Santos},
  {Giard}, {Gilfanov}, {Giraud-H{\'e}raud}, {Gonz{\'a}lez-Nuevo}, {G{\'o}rski},
  {Gregorio}, {Gruppuso}, {Hansen}, {Harrison}, {Hempel},
  {Henrot-Versill{\'e}}, {Hern{\'a}ndez-Monteagudo}, {Herranz}, {Hildebrandt},
  {Hivon}, {Hobson}, {Holmes}, {Hovest}, {Hurier}, {Jaffe}, {Jaffe},
  {Jagemann}, {Jones}, {Juvela}, {Khamitov}, {Kisner}, {Kneissl}, {Knoche},
  {Knox}, {Kunz}, {Kurki-Suonio}, {Lagache}, {Lamarre}, {Lasenby}, {Lawrence},
  {Le Jeune}, {Leonardi}, {Lilje}, {Linden-V{\o}rnle}, {L{\'o}pez-Caniego},
  {Lubin}, {Luzzi}, {Mac{\'\i}as-P{\'e}rez}, {Maffei}, {Maino}, {Mandolesi},
  {Maris}, {Marleau}, {Marshall}, {Mart{\'\i}nez-Gonz{\'a}lez}, {Masi},
  {Massardi}, {Matarrese}, {Matthai}, {Mazzotta}, {Mei}, {Melchiorri}, {Melin},
  {Mendes}, {Mennella}, {Mitra}, {Miville-Desch{\`e}nes}, {Moneti}, {Montier},
  {Morgante}, {Munshi}, {Murphy}, {Naselsky}, {Nati}, {Natoli},
  {N{\o}rgaard-Nielsen}, {Noviello}, {Novikov}, {Novikov}, {Osborne}, {Pajot},
  {Paoletti}, {Pasian}, {Patanchon}, {Perdereau}, {Perotto}, {Perrotta},
  {Piacentini}, {Piat}, {Pierpaoli}, {Piffaretti}, {Plaszczynski},
  {Pointecouteau}, {Polenta}, {Ponthieu}, {Popa}, {Poutanen}, {Pratt},
  {Prunet}, {Puget}, {Rachen}, {Rebolo}, {Reinecke}, {Remazeilles}, {Renault},
  {Ricciardi}, {Riller}, {Ristorcelli}, {Rocha}, {Roman}, {Rosset}, {Rossetti},
  {Rubi{\~n}o-Mart{\'\i}n}, {Rusholme}, {Sandri}, {Savini}, {Schaefer},
  {Scott}, {Smoot}, {Starck}, {Sudiwala}, {Sunyaev}, {Sutton}, {Suur-Uski},
  {Sygnet}, {Tauber}, {Terenzi}, {Toffolatti}, {Tomasi}, {Tristram}, {Tucci},
  {Valenziano}, {Van Tent}, {Vielva}, {Villa}, {Vittorio}, {Wade}, {Wandelt},
  {Welikala}, {White}, {Yvon}, {Zacchei}, \& {Zonca}}]{Planck2013_filament}
{Planck Collaboration}, {Ade}, P.~A.~R., {Aghanim}, N., {et~al.} 2013, \aap,
  550, A134

\bibitem[{{Planck Collaboration} {et~al.}(2016){Planck Collaboration}, {Ade},
  {Aghanim}, {Arnaud}, {Ashdown}, {Aumont}, {Baccigalupi}, {Banday},
  {Barreiro}, {Bartlett}, {Bartolo}, {Battaner}, {Battye}, {Benabed},
  {Beno{\^\i}t}, {Benoit-L{\'e}vy}, {Bernard}, {Bersanelli}, {Bielewicz},
  {Bock}, {Bonaldi}, {Bonavera}, {Bond}, {Borrill}, {Bouchet}, {Bucher},
  {Burigana}, {Butler}, {Calabrese}, {Cardoso}, {Catalano}, {Challinor},
  {Chamballu}, {Chary}, {Chiang}, {Christensen}, {Church}, {Clements},
  {Colombi}, {Colombo}, {Combet}, {Comis}, {Couchot}, {Coulais}, {Crill},
  {Curto}, {Cuttaia}, {Danese}, {Davies}, {Davis}, {de Bernardis}, {de Rosa},
  {de Zotti}, {Delabrouille}, {D{\'e}sert}, {Diego}, {Dolag}, {Dole},
  {Donzelli}, {Dor{\'e}}, {Douspis}, {Ducout}, {Dupac}, {Efstathiou}, {Elsner},
  {En{\ss}lin}, {Eriksen}, {Falgarone}, {Fergusson}, {Finelli}, {Forni},
  {Frailis}, {Fraisse}, {Franceschi}, {Frejsel}, {Galeotta}, {Galli}, {Ganga},
  {Giard}, {Giraud-H{\'e}raud}, {Gjerl{\o}w}, {Gonz{\'a}lez-Nuevo},
  {G{\'o}rski}, {Gratton}, {Gregorio}, {Gruppuso}, {Gudmundsson}, {Hansen},
  {Hanson}, {Harrison}, {Henrot-Versill{\'e}}, {Hern{\'a}ndez-Monteagudo},
  {Herranz}, {Hildebrandt}, {Hivon}, {Hobson}, {Holmes}, {Hornstrup}, {Hovest},
  {Huffenberger}, {Hurier}, {Jaffe}, {Jaffe}, {Jones}, {Juvela},
  {Keih{\"a}nen}, {Keskitalo}, {Kisner}, {Kneissl}, {Knoche}, {Kunz},
  {Kurki-Suonio}, {Lagache}, {L{\"a}hteenm{\"a}ki}, {Lamarre}, {Lasenby},
  {Lattanzi}, {Lawrence}, {Leonardi}, {Lesgourgues}, {Levrier}, {Liguori},
  {Lilje}, {Linden-V{\o}rnle}, {L{\'o}pez-Caniego}, {Lubin},
  {Mac{\'\i}as-P{\'e}rez}, {Maggio}, {Maino}, {Mand olesi}, {Mangilli},
  {Maris}, {Martin}, {Mart{\'\i}nez-Gonz{\'a}lez}, {Masi}, {Matarrese},
  {McGehee}, {Meinhold}, {Melchiorri}, {Melin}, {Mendes}, {Mennella},
  {Migliaccio}, {Mitra}, {Miville-Desch{\^e}nes}, {Moneti}, {Montier},
  {Morgante}, {Mortlock}, {Moss}, {Munshi}, {Murphy}, {Naselsky}, {Nati},
  {Natoli}, {Netterfield}, {N{\o}rgaard-Nielsen}, {Noviello}, {Novikov},
  {Novikov}, {Oxborrow}, {Paci}, {Pagano}, {Pajot}, {Paoletti}, {Partridge},
  {Pasian}, {Patanchon}, {Pearson}, {Perdereau}, {Perotto}, {Perrotta},
  {Pettorino}, {Piacentini}, {Piat}, {Pierpaoli}, {Pietrobon}, {Plaszczynski},
  {Pointecouteau}, {Polenta}, {Popa}, {Pratt}, {Pr{\'e}zeau}, {Prunet},
  {Puget}, {Rachen}, {Rebolo}, {Reinecke}, {Remazeilles}, {Renault}, {Renzi},
  {Ristorcelli}, {Rocha}, {Roman}, {Rosset}, {Rossetti}, {Roudier},
  {Rubi{\~n}o-Mart{\'\i}n}, {Rusholme}, {Sandri}, {Santos}, {Savelainen},
  {Savini}, {Scott}, {Seiffert}, {Shellard}, {Spencer}, {Stolyarov}, {Stompor},
  {Sudiwala}, {Sunyaev}, {Sutton}, {Suur-Uski}, {Sygnet}, {Tauber}, {Terenzi},
  {Toffolatti}, {Tomasi}, {Tristram}, {Tucci}, {Tuovinen}, {T{\"u}rler},
  {Umana}, {Valenziano}, {Valiviita}, {Van Tent}, {Vielva}, {Villa}, {Wade},
  {Wandelt}, {Wehus}, {Weller}, {White}, {Yvon}, {Zacchei}, \&
  {Zonca}}]{Planck2016}
{Planck Collaboration}, {Ade}, P.~A.~R., {Aghanim}, N., {et~al.} 2016, \aap,
  594, A24

\bibitem[{{Power} {et~al.}(2012){Power}, {Knebe}, \& {Knollmann}}]{Power2012}
{Power}, C., {Knebe}, A., \& {Knollmann}, S.~R. 2012, \mnras, 419, 1576

\bibitem[{{Predehl} {et~al.}(2021){Predehl}, {Andritschke}, {Arefiev},
  {Babyshkin}, {Batanov}, {Becker}, {B{\"o}hringer}, {Bogomolov}, {Boller},
  {Borm}, {Bornemann}, {Br{\"a}uninger}, {Br{\"u}ggen}, {Brunner}, {Brusa},
  {Bulbul}, {Buntov}, {Burwitz}, {Burkert}, {Clerc}, {Churazov}, {Coutinho},
  {Dauser}, {Dennerl}, {Doroshenko}, {Eder}, {Emberger}, {Eraerds},
  {Finoguenov}, {Freyberg}, {Friedrich}, {Friedrich}, {F{\"u}rmetz},
  {Georgakakis}, {Gilfanov}, {Granato}, {Grossberger}, {Gueguen}, {Gureev},
  {Haberl}, {H{\"a}lker}, {Hartner}, {Hasinger}, {Huber}, {Ji}, {Kienlin},
  {Kink}, {Korotkov}, {Kreykenbohm}, {Lamer}, {Lomakin}, {Lapshov}, {Liu},
  {Maitra}, {Meidinger}, {Menz}, {Merloni}, {Mernik}, {Mican}, {Mohr},
  {M{\"u}ller}, {Nandra}, {Nazarov}, {Pacaud}, {Pavlinsky}, {Perinati},
  {Pfeffermann}, {Pietschner}, {Ramos-Ceja}, {Rau}, {Reiffers}, {Reiprich},
  {Robrade}, {Salvato}, {Sanders}, {Santangelo}, {Sasaki}, {Scheuerle},
  {Schmid}, {Schmitt}, {Schwope}, {Shirshakov}, {Steinmetz}, {Stewart},
  {Str{\"u}der}, {Sunyaev}, {Tenzer}, {Tiedemann}, {Tr{\"u}mper}, {Voron},
  {Weber}, {Wilms}, \& {Yaroshenko}}]{eRosita}
{Predehl}, P., {Andritschke}, R., {Arefiev}, V., {et~al.} 2021, \aap, 647, A1

\bibitem[{{Reiprich} {et~al.}(2021){Reiprich}, {Veronica}, {Pacaud},
  {Ramos-Ceja}, {Ota}, {Sanders}, {Kara}, {Erben}, {Klein}, {Erler}, {Kerp},
  {Hoang}, {Br{\"u}ggen}, {Marvil}, {Rudnick}, {Biffi}, {Dolag},
  {Aschersleben}, {Basu}, {Brunner}, {Bulbul}, {Dennerl}, {Eckert}, {Freyberg},
  {Gatuzz}, {Ghirardini}, {K{\"a}fer}, {Merloni}, {Migkas}, {Nandra},
  {Predehl}, {Robrade}, {Salvato}, {Whelan}, {Diaz-Ocampo}, {Hernandez-Lang},
  {Zenteno}, {Brown}, {Collier}, {Diego}, {Hopkins}, {Kapinska}, {Koribalski},
  {Mroczkowski}, {Norris}, {O'Brien}, \& {Vardoulaki}}]{EROSITA_Reiprich2021}
{Reiprich}, T.~H., {Veronica}, A., {Pacaud}, F., {et~al.} 2021, \aap, 647, A2

\bibitem[{{Rodriguez-Gomez} {et~al.}(2015){Rodriguez-Gomez}, {Genel},
  {Vogelsberger}, {Sijacki}, {Pillepich}, {Sales}, {Torrey}, {Snyder},
  {Nelson}, {Springel}, {Ma}, \& {Hernquist}}]{Rodriguez2015}
{Rodriguez-Gomez}, V., {Genel}, S., {Vogelsberger}, M., {et~al.} 2015, \mnras,
  449, 49

\bibitem[{{Roncarelli} {et~al.}(2013){Roncarelli}, {Ettori}, {Borgani},
  {Dolag}, {Fabjan}, \& {Moscardini}}]{Roncarelli2013}
{Roncarelli}, M., {Ettori}, S., {Borgani}, S., {et~al.} 2013, \mnras, 432, 3030

\bibitem[{{Rost} {et~al.}(2021){Rost}, {Kuchner}, {Welker}, {Pearce},
  {Stasyszyn}, {Gray}, {Cui}, {Dave}, {Knebe}, {Yepes}, \& {Rasia}}]{Rost2021}
{Rost}, A., {Kuchner}, U., {Welker}, C., {et~al.} 2021, \mnras, 502, 714

\bibitem[{{Santos} {et~al.}(2008){Santos}, {Rosati}, {Tozzi}, {B{\"o}hringer},
  {Ettori}, \& {Bignamini}}]{Santos2008}
{Santos}, J.~S., {Rosati}, P., {Tozzi}, P., {et~al.} 2008, \aap, 483, 35

\bibitem[{{Sarron} {et~al.}(2019){Sarron}, {Adami}, {Durret}, \&
  {Laigle}}]{Sarron2019}
{Sarron}, F., {Adami}, C., {Durret}, F., \& {Laigle}, C. 2019, \aap, 632, A49

\bibitem[{{Schade} {et~al.}(1995){Schade}, {Lilly}, {Crampton}, {Hammer}, {Le
  Fevre}, \& {Tresse}}]{Schade1995}
{Schade}, D., {Lilly}, S.~J., {Crampton}, D., {et~al.} 1995, \apjl, 451, L1

\bibitem[{{Schneider} \& {Bartelmann}(1997)}]{Schneider1997}
{Schneider}, P. \& {Bartelmann}, M. 1997, \mnras, 286, 696

\bibitem[{{Schneider} \& {Weiss}(1991)}]{Schneider1991}
{Schneider}, P. \& {Weiss}, A. 1991, \aap, 247, 269

\bibitem[{{Sembolini} {et~al.}(2016){Sembolini}, {Elahi}, {Pearce}, {Power},
  {Knebe}, {Kay}, {Cui}, {Yepes}, {Beck}, {Borgani}, {Cunnama}, {Dav{\'e}},
  {February}, {Huang}, {Katz}, {McCarthy}, {Murante}, {Newton}, {Perret},
  {Puchwein}, {Saro}, {Schaye}, \& {Teyssier}}]{Sembolini2016}
{Sembolini}, F., {Elahi}, P.~J., {Pearce}, F.~R., {et~al.} 2016, \mnras, 459,
  2973

\bibitem[{{Shi} {et~al.}(2020){Shi}, {Nagai}, {Aung}, \& {Wetzel}}]{Shi2020}
{Shi}, X., {Nagai}, D., {Aung}, H., \& {Wetzel}, A. 2020, \mnras, 495, 784

\bibitem[{{Shim} {et~al.}(2020){Shim}, {Codis}, {Pichon}, {Pogosyan}, \&
  {Cadiou}}]{Shim2020}
{Shim}, J., {Codis}, S., {Pichon}, C., {Pogosyan}, D., \& {Cadiou}, C. 2020,
  arXiv e-prints, arXiv:2011.04321

\bibitem[{{Shin} {et~al.}(2018){Shin}, {Clampitt}, {Jain}, {Bernstein}, {Neil},
  {Rozo}, \& {Rykoff}}]{Shin2018}
{Shin}, T.-h., {Clampitt}, J., {Jain}, B., {et~al.} 2018, \mnras, 475, 2421

\bibitem[{{Simionescu} {et~al.}(2021){Simionescu}, {Ettori}, {Werner}, {Nagai},
  {Vazza}, {Akamatsu}, {Pinto}, {de Plaa}, {Wijers}, {Nelson}, {Pointecouteau},
  {Pratt}, {Spiga}, {Vacanti}, {Lau}, {Rossetti}, {Gastaldello}, {Biffi},
  {Bulbul}, {Collon}, {Herder}, {Eckert}, {Fraternali}, {Mingo}, {Pareschi},
  {Pezzulli}, {Reiprich}, {Schaye}, {Walker}, \& {Werk}}]{Simionescu2021}
{Simionescu}, A., {Ettori}, S., {Werner}, N., {et~al.} 2021, Experimental
  Astronomy

\bibitem[{{Singh} {et~al.}(2020){Singh}, {Mahajan}, \& {Bagla}}]{Singh2020}
{Singh}, A., {Mahajan}, S., \& {Bagla}, J.~S. 2020, \mnras, 497, 2265

\bibitem[{{Song} {et~al.}(2021){Song}, {Laigle}, {Hwang}, {Devriendt},
  {Dubois}, {Kraljic}, {Pichon}, {Slyz}, \& {Smith}}]{Song2021}
{Song}, H., {Laigle}, C., {Hwang}, H.~S., {et~al.} 2021, \mnras, 501, 4635

\bibitem[{{Sorini} {et~al.}(2021){Sorini}, {Dave}, {Cui}, \&
  {Appleby}}]{Sorini2021}
{Sorini}, D., {Dave}, R., {Cui}, W., \& {Appleby}, S. 2021, arXiv e-prints,
  arXiv:2111.13708

\bibitem[{{Sousbie}(2011)}]{Sousbie2011}
{Sousbie}, T. 2011, \mnras, 414, 350

\bibitem[{{Springel}(2010)}]{AREPO}
{Springel}, V. 2010, \mnras, 401, 791

\bibitem[{{Springel} {et~al.}(2001){Springel}, {White}, {Tormen}, \&
  {Kauffmann}}]{subfind}
{Springel}, V., {White}, S. D.~M., {Tormen}, G., \& {Kauffmann}, G. 2001,
  \mnras, 328, 726

\bibitem[{{Suto} {et~al.}(2016){Suto}, {Kitayama}, {Nishimichi}, {Sasaki}, \&
  {Suto}}]{Suto2016}
{Suto}, D., {Kitayama}, T., {Nishimichi}, T., {Sasaki}, S., \& {Suto}, Y. 2016,
  \pasj, 68, 97

\bibitem[{{Tanimura} {et~al.}(2020{\natexlab{a}}){Tanimura}, {Aghanim},
  {Bonjean}, {Malavasi}, \& {Douspis}}]{Tanimura2020A}
{Tanimura}, H., {Aghanim}, N., {Bonjean}, V., {Malavasi}, N., \& {Douspis}, M.
  2020{\natexlab{a}}, \aap, 637, A41

\bibitem[{{Tanimura} {et~al.}(2020{\natexlab{b}}){Tanimura}, {Aghanim},
  {Kolodzig}, {Douspis}, \& {Malavasi}}]{Tanimura2020B}
{Tanimura}, H., {Aghanim}, N., {Kolodzig}, A., {Douspis}, M., \& {Malavasi}, N.
  2020{\natexlab{b}}, \aap, 643, L2

\bibitem[{{Tanimura} {et~al.}(2019){Tanimura}, {Hinshaw}, {McCarthy}, {Van
  Waerbeke}, {Aghanim}, {Ma}, {Mead}, {Hojjati}, \&
  {Tr{\"o}ster}}]{Tanimura2019}
{Tanimura}, H., {Hinshaw}, G., {McCarthy}, I.~G., {et~al.} 2019, \mnras, 483,
  223

\bibitem[{{Tchernin} {et~al.}(2016){Tchernin}, {Eckert}, {Ettori},
  {Pointecouteau}, {Paltani}, {Molendi}, {Hurier}, {Gastaldello}, {Lau},
  {Nagai}, {Roncarelli}, \& {Rossetti}}]{XCOP}
{Tchernin}, C., {Eckert}, D., {Ettori}, S., {et~al.} 2016, \aap, 595, A42

\bibitem[{{Tempel} {et~al.}(2016){Tempel}, {Stoica}, {Kipper}, \&
  {Saar}}]{Tempel2016}
{Tempel}, E., {Stoica}, R.~S., {Kipper}, R., \& {Saar}, E. 2016, Astronomy and
  Computing, 16, 17

\bibitem[{{Truong} {et~al.}(2021){Truong}, {Pillepich}, \&
  {Werner}}]{Truong2021}
{Truong}, N., {Pillepich}, A., \& {Werner}, N. 2021, \mnras, 501, 2210

\bibitem[{{Tuominen} {et~al.}(2021){Tuominen}, {Nevalainen}, {Tempel},
  {Kuutma}, {Wijers}, {Schaye}, {Hein{\"a}m{\"a}ki}, {Bonamente}, \&
  {Ganeshaiah Veena}}]{Tuominen2021}
{Tuominen}, T., {Nevalainen}, J., {Tempel}, E., {et~al.} 2021, \aap, 646, A156

\bibitem[{{Vall{\'e}s-P{\'e}rez} {et~al.}(2020){Vall{\'e}s-P{\'e}rez},
  {Planelles}, \& {Quilis}}]{Valles2020}
{Vall{\'e}s-P{\'e}rez}, D., {Planelles}, S., \& {Quilis}, V. 2020, \mnras, 499,
  2303

\bibitem[{{Vall{\'e}s-P{\'e}rez} {et~al.}(2021){Vall{\'e}s-P{\'e}rez},
  {Planelles}, \& {Quilis}}]{Valles2021}
{Vall{\'e}s-P{\'e}rez}, D., {Planelles}, S., \& {Quilis}, V. 2021, \mnras, 504,
  510

\bibitem[{{Vazza} {et~al.}(2011){Vazza}, {Roncarelli}, {Ettori}, \&
  {Dolag}}]{Vazza2011}
{Vazza}, F., {Roncarelli}, M., {Ettori}, S., \& {Dolag}, K. 2011, \mnras, 413,
  2305

\bibitem[{{Velliscig} {et~al.}(2015){Velliscig}, {Cacciato}, {Schaye}, {Crain},
  {Bower}, {van Daalen}, {Dalla Vecchia}, {Frenk}, {Furlong}, {McCarthy},
  {Schaller}, \& {Theuns}}]{Velliscig2015}
{Velliscig}, M., {Cacciato}, M., {Schaye}, J., {et~al.} 2015, \mnras, 453, 721

\bibitem[{{Veronica} {et~al.}(2021){Veronica}, {Su}, {Biffi}, {Reiprich},
  {Pacaud}, {Nulsen}, {Kraft}, {Sanders}, {Bogdan}, {Kara}, {Dolag}, {Kerp},
  {Koribalski}, {Erben}, {Bulbul}, {Gatuzz}, {Ghirardini}, {Hopkins}, {Liu},
  {Migkas}, \& {Vernstrom}}]{EROSITA_Angie2021}
{Veronica}, A., {Su}, Y., {Biffi}, V., {et~al.} 2021, arXiv e-prints,
  arXiv:2106.14543

\bibitem[{{Walker} {et~al.}(2019){Walker}, {Simionescu}, {Nagai}, {Okabe},
  {Eckert}, {Mroczkowski}, {Akamatsu}, {Ettori}, \& {Ghirardini}}]{Walker2019}
{Walker}, S., {Simionescu}, A., {Nagai}, D., {et~al.} 2019, \ssr, 215, 7

\bibitem[{{XRISM Science Team}(2020)}]{XRISM}
{XRISM Science Team}. 2020, arXiv e-prints, arXiv:2003.04962

\bibitem[{{Zel'Dovich}(1970)}]{Zeldovich1970}
{Zel'Dovich}, Y.~B. 1970, \aap, 500, 13

\bibitem[{{Zhang} {et~al.}(2021{\natexlab{a}}){Zhang}, {Cui}, {Dave}, \& {De
  Petris}}]{Zhang2021_300_DS}
{Zhang}, B., {Cui}, W., {Dave}, R., \& {De Petris}, M. 2021{\natexlab{a}},
  arXiv e-prints, arXiv:2112.01909

\bibitem[{{Zhang} {et~al.}(2020){Zhang}, {Churazov}, {Dolag}, {Forman}, \&
  {Zhuravleva}}]{Zhang2020}
{Zhang}, C., {Churazov}, E., {Dolag}, K., {Forman}, W.~R., \& {Zhuravleva}, I.
  2020, \mnras, 494, 4539

\bibitem[{{Zhang} {et~al.}(2021{\natexlab{b}}){Zhang}, {Zhuravleva},
  {Kravtsov}, \& {Churazov}}]{Zhang2021}
{Zhang}, C., {Zhuravleva}, I., {Kravtsov}, A., \& {Churazov}, E.
  2021{\natexlab{b}}, \mnras, 506, 839

\bibitem[{{Zhu} {et~al.}(2021{\natexlab{a}}){Zhu}, {Zhang}, \&
  {Feng}}]{Zhu2021}
{Zhu}, W., {Zhang}, F., \& {Feng}, L.-L. 2021{\natexlab{a}}, arXiv e-prints,
  arXiv:2107.08663

\bibitem[{{Zhu} {et~al.}(2021{\natexlab{b}}){Zhu}, {Simionescu}, {Akamatsu},
  {Zhang}, {Kaastra}, {de Plaa}, {Urban}, {Allen}, \& {Werner}}]{Zhu2021_shock}
{Zhu}, Z., {Simionescu}, A., {Akamatsu}, H., {et~al.} 2021{\natexlab{b}}, \aap,
  652, A147

\bibitem[{{Zhuravleva} {et~al.}(2013){Zhuravleva}, {Churazov}, {Kravtsov},
  {Lau}, {Nagai}, \& {Sunyaev}}]{Zhuravleva2013}
{Zhuravleva}, I., {Churazov}, E., {Kravtsov}, A., {et~al.} 2013, \mnras, 428,
  3274

\end{thebibliography}

\appendix

\section{Multipolar expansion of DM distribution\label{APPENDIX_BETA}}

We provide here complementary information on multipolar decomposition of 2-D mass distribution in harmonic space.
As presented in \cite{Schneider1997,Gouin2017,Codis2017,Gouin2020}, matter distribution can be decomposed in different harmonic modes $m$ to compute multipole moment $Q_m$ (see Eq. \ref{eq:multipole_qm}). We illustrate the different azimuthal symmetries in Fig. \ref{fig:Annex_0} as a function of the orders $m$, such that one can see the monopole ($m=0$), dipole ($m=1$), quadrupole ($m=2$), etc.

\begin{figure}
    \centering
    \includegraphics[width=0.48\textwidth]{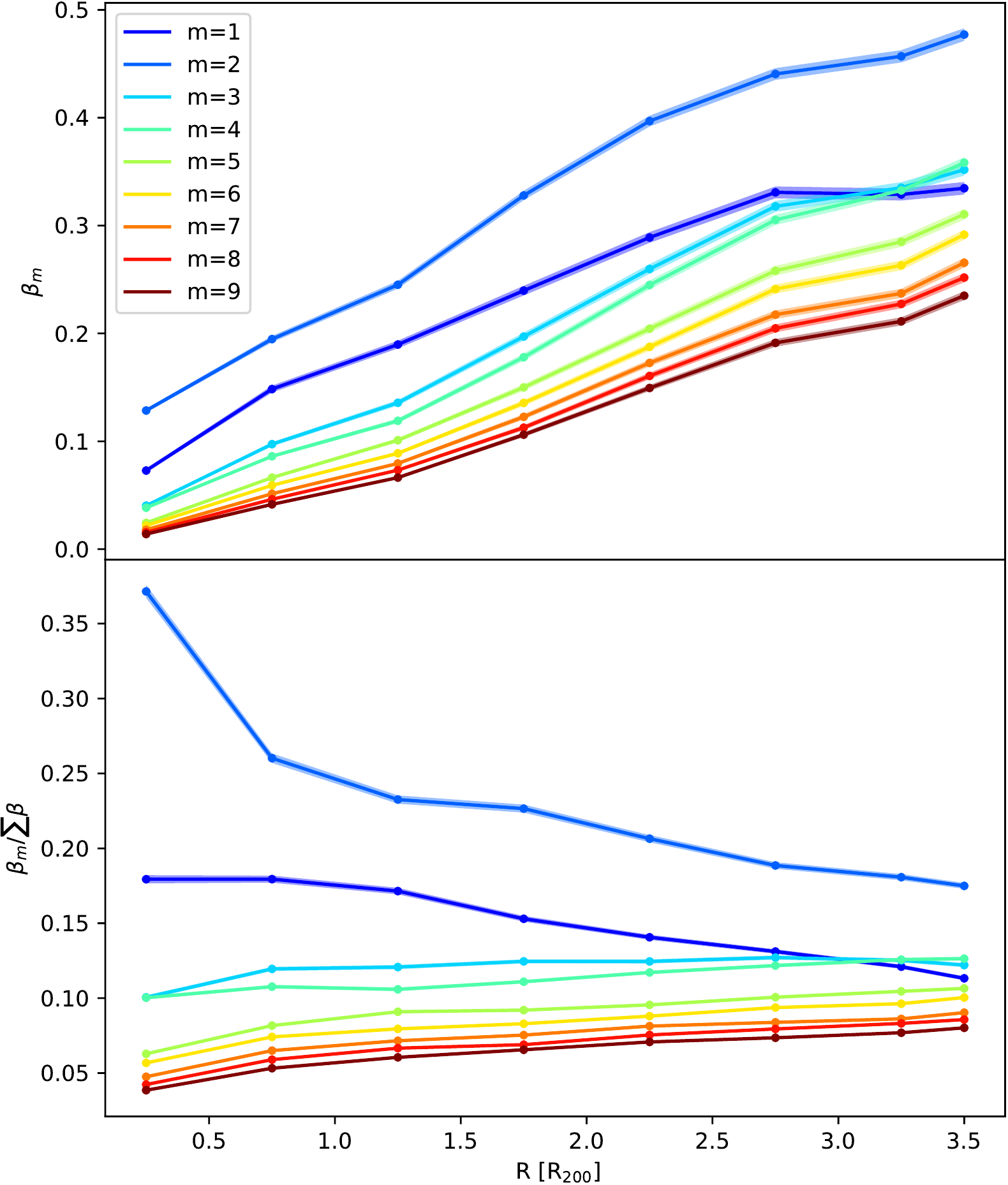}
    \caption{Top panel: Mean evolution of $\beta_m$ parameter for $m=1$ to $m=9$ as a function of the cluster radial distance. 
    Bottom panel: Mean evolution of $\beta_m$ parameter normalised by the sum of all order contributions $\sum_{i=1}^{9} \beta_i$ from $m=1$ to $m=9$, as a function of the cluster radial distance. 
    \label{fig:Annex_1}}
\end{figure}

To confirm the previous findings form \cite{Gouin2020} and \cite{Valles2020}, we plot Fig. \ref{fig:Annex_1} the mean evolution of multipolar ratio $\beta_m$ for $m=1$ to $m=9$ as a function of the cluster radial distance (top panel). 
First, one can see that the level of azimuthal symmetry traced by $\beta_m$ increases with the radial distance for all orders $m$. In agreement with the azimuthal scatter technique, departure from spherical symmetries increases as a function of the radial-cluster distance, as measured in observations and hydrodynamical simulations \citep[see the figure 7 and 9 of][]{Eckert2012}.
In the bottom panel of Fig. \ref{fig:Annex_1}, we show that the dominant azimuthal order is the quadrupole $m=2$ \citep[see also][]{Valles2020} and the second most significant azimuthal symmetric orders are $m=1,3,4$ with contributions higher than 10\% \citep[see also][]{Gouin2020}. In detail, the azimuthal matter distribution inside galaxy clusters (up to $R_{200}$) is almost quadrupolar, whereas the  matter distribution in cluster infalling regions (from $R_{200}$ to $4 \times R_{200}$) is mainly traced by 
the harmonic orders $m=1,2,3,4$. At these distances from cluster centers, matter infalls from cosmic filaments that are connected to clusters \citep{Rost2021}.
Therefore, in Sect. \ref{SEC:MAH}, we focus on these orders to analyse the relation between azimuthal matter distribution and cluster properties.

\section{Quadrupolar signature and ellipticity \label{APPENDIX_ELLIPSE}}

\begin{figure}
    \centering
    \includegraphics[width=0.3\textwidth]{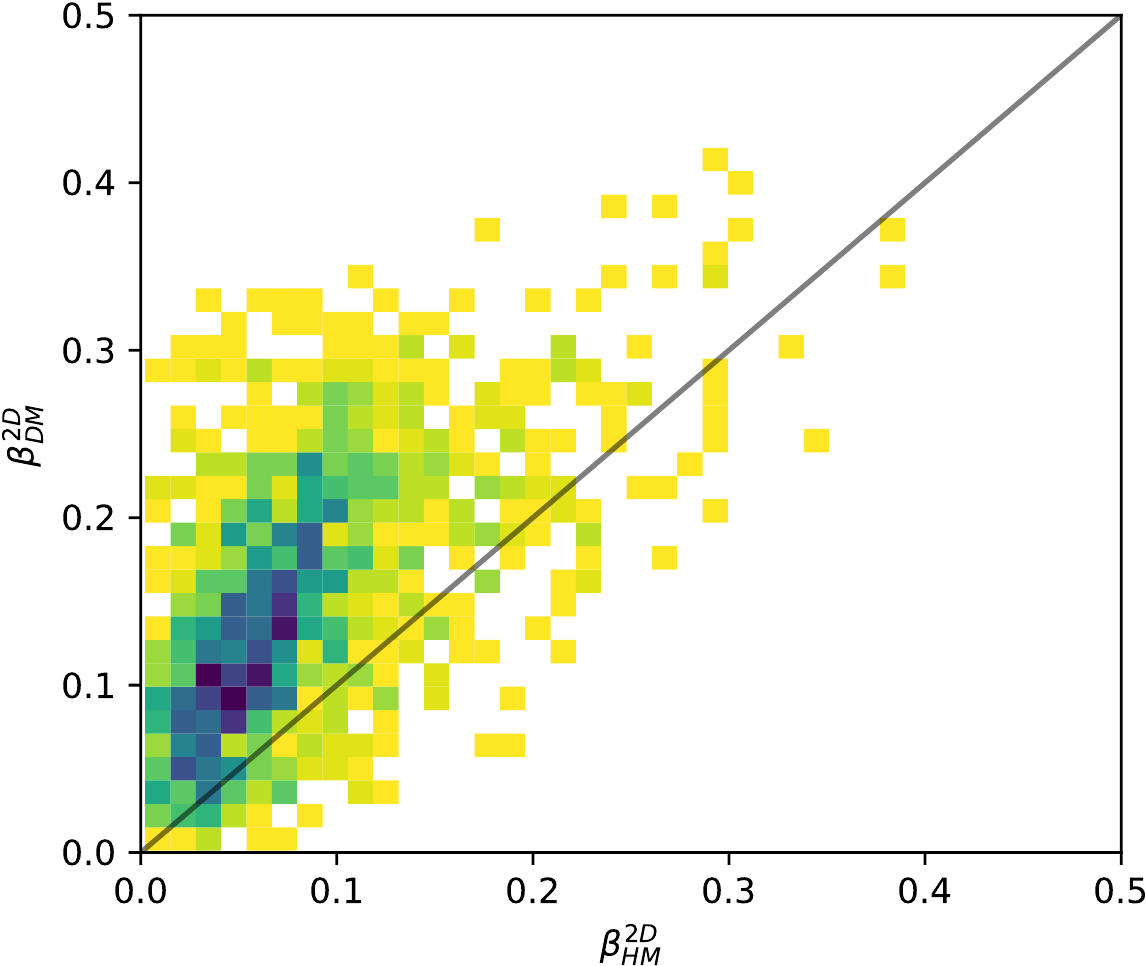}
    \includegraphics[width=0.3\textwidth]{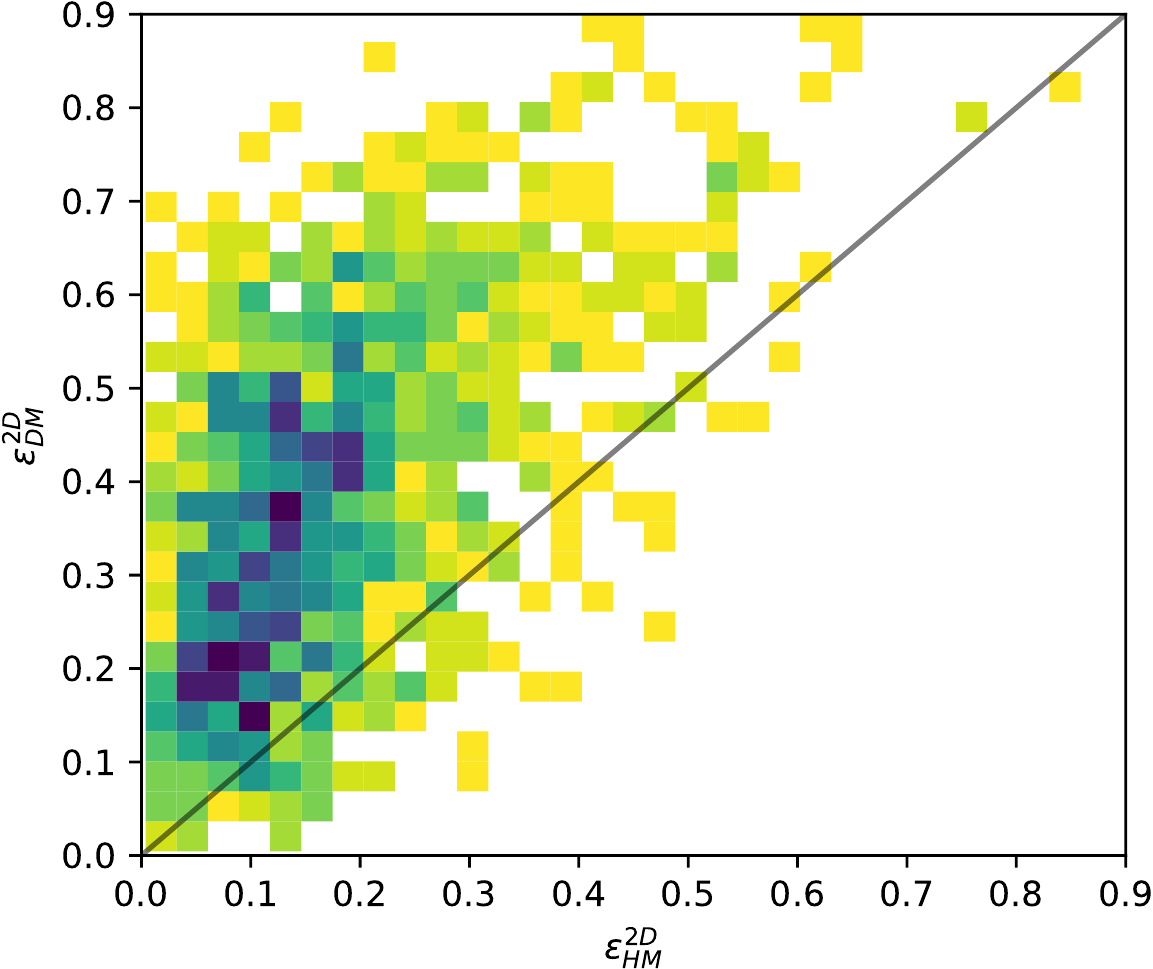}
    \caption{Distribution of the quadrupolar ratio $\beta_2$ (top panel) and 2-D ellipticity $\epsilon$ (bottom panel) of dark matter and hot gas for our all cluster sample. \label{fig:Annex_ellipticity} }
\end{figure}

As shown in the top right panel in Fig. \ref{fig:FIG_STRUCURAL}, the quadrupole symmetry $\beta_2$ can be related to the ellipticity $\epsilon$.
We illustrate in Fig. \ref{fig:Annex_ellipticity} the distribution of quadrupolar ratios of hot gas and DM for all halos in top panel, and the distribution of their 2-D ellipticities in bottom panel. Both show similar behavior, and thus, confirm that hot gas is more circular (less elliptical) than the DM distribution on average, in agreement with \cite{Velliscig2015,Donahue2016,Okabe2018}. Therefore, the quadrupolar ratio appears to be a good tracer of the ellipticity for both DM and hot gas components inside clusters.

\end{document}